\documentclass[12pt,letterpaper]{article}
\pdfoutput=1
\usepackage{graphicx,array}
\usepackage{color}
\usepackage{latexsym}
\usepackage{amsthm}
\usepackage{amsmath}
\usepackage{amssymb}
\usepackage{hyperref}
\usepackage{bbold}
\usepackage{subfig}

\def\simgt{\mathrel{\lower2.5pt\vbox{\lineskip=0pt\baselineskip=0pt
           \hbox{$>$}\hbox{$\sim$}}}}
\def\simlt{\mathrel{\lower2.5pt\vbox{\lineskip=0pt\baselineskip=0pt
           \hbox{$<$}\hbox{$\sim$}}}}

\newcommand{\be}{\begin{equation}}
\newcommand{\ee}{\end{equation}}
\newcommand{\bea}{\begin{eqnarray}}
\newcommand{\eea}{\end{eqnarray}}
\newcommand{\Eq}[1]{Eq.~(\ref{#1})}

\newcommand{\Fig}[1]{Fig.~(\ref{#1})}

\newcommand{\vev}[1]{\langle #1 \rangle}
\newcommand{\GeV}{\textrm{ GeV}}
\newcommand{\TeV}{\textrm{ TeV}}
\newcommand{\gsim}{\lower.7ex\hbox{$\;\stackrel{\textstyle>}{\sim}\;$}}
\newcommand{\lsim}{\lower.7ex\hbox{$\;\stackrel{\textstyle<}{\sim}\;$}}

\hypersetup{colorlinks,citecolor= blue,linkcolor= black}

\begin{document}


\hfill

\vspace{1.5cm}

\begin{center}
{\LARGE\bf
Radiative PQ Breaking \\ and  \vspace{.4cm}\\
The Higgs Boson Mass 
}
\\ \vspace*{0.5cm}

\bigskip\vspace{1cm}{
{\large Francesco D'Eramo, Lawrence J. Hall and Duccio Pappadopulo}
} \\[7mm]
 {\it 

 Berkeley Center for Theoretical Physics, Department of Physics, \\
     and Theoretical Physics Group, Lawrence Berkeley National Laboratory, \\
     University of California, Berkeley, CA 94720, USA} \end{center}
\bigskip
\centerline{\large\bf Abstract}

\vspace{0.3cm}

The small and negative value of the Standard Model Higgs quartic coupling at high scales can be understood in terms of anthropic selection on a landscape where large and negative values are favored: most universes have a very short-lived electroweak vacuum and typical observers are in universes close to the corresponding metastability boundary. We provide a simple example of such a landscape with a Peccei-Quinn symmetry breaking scale generated through dimensional transmutation and supersymmetry softly broken at an intermediate scale.  Large and negative contributions to the Higgs quartic are typically generated on integrating out the saxion field.  Cancellations among these contributions are forced by the anthropic requirement of a sufficiently long-lived electroweak vacuum, determining the multiverse distribution for the Higgs quartic in a similar way to that of the cosmological constant.  This leads to a  statistical prediction of the Higgs boson mass that, for a wide range of parameters, yields the observed value within the $1 \sigma$ statistical uncertainty of $\sim 5$ GeV originating from the multiverse distribution. The strong CP problem is solved and single-component axion dark matter is predicted, with an abundance that can be understood from environmental selection.  A more general setting for the Higgs mass prediction is discussed.

\begin{quote} \small

\end{quote}

\newpage
\tableofcontents
\newpage

\section{Introduction}
\label{sec:intro}
The discovery of a perturbative Higgs at LHC, together with the absence of any signal in recent decades of new physics at particle accelerators or in searches for dark matter, suggests the possibility of a paradigm shift in Beyond Standard Model Physics:  the Standard Model (SM) may be valid up to scales very much larger than the weak scale.  The key questions then become very different from those associated with new physics at the TeV scale, and in this paper we focus on 
\begin{itemize}
\item The origin of an unnatural weak scale, $v$.
\item The origin for an unnaturally small value for CP violation in QCD, $\bar{\theta}$.
\item The origin for the small Higgs quartic coupling at very high scales, $\lambda$.  
\item The nature and cosmological abundance of dark matter, $\rho_m$.
\end{itemize}
The multiverse provides a framework for understanding the highly fine-tuned values of both the cosmological constant \cite{Weinberg:1987dv,Martel:1997vi}, including a solution to the Why Now Problem \cite{Bousso:2007kq, Bousso:2010zi, Bousso:2010im}, and the weak scale \cite{Agrawal:1997gf, Hall:2014dfa}.    However, it does not explain the smallness of $\bar{\theta}$.  By far the most compelling understanding of $\bar{\theta}$ is provided by a Peccei-Quinn symmetry \cite{Peccei:1977hh}, which promotes $\bar{\theta}$ to a field.  Furthermore the resulting axion  \cite{Weinberg:1977ma, Wilczek:1977pj} can account for the observed dark matter.  

Can the multiverse provide an explanation for the $\lambda$?  In this paper we provide a particularly simple model for the axion that relies heavily on the multiverse, not just to account for the observed value of $v$ and perhaps $\rho_m$, but because it typically leads to a large negative value for the Higgs quartic coupling $\lambda$ at high scales.  From a conventional view, where $v, \rho_m, \lambda$ are to be understood from symmetries, this model is disastrous.   But in the multiverse view, parameters that determine the Higgs potential must be anthropically selected to yield a universe with a sufficiently stable electroweak vacuum.  This allows us to compute the probability distribution for $\lambda$ and obtain a successful statistical prediction for the Higgs boson mass.  

We study a supersymmetric extension of the SM that has a PQ symmetry and is valid to some very high energy scale $M_*$ that may be near the Planck scale.   Below $M_*$ the theory contains no dimensionful parameters other than the scale $\tilde{m}$ of supersymmetry breaking, which we take to scan.  The Higgs sector of the theory contains a gauge singlet field $S$ in addition to two Higgs doublets with the interaction
\be
W_S \; = \; \xi \, S H_u H_d 
\label{eq:W}
\ee
where $\xi$ is an order unity coupling.  One possibility is that this theory is the MSSM with $\mu$ replaced by $S$.  This is {\it not} the NMSSM since $S$ is charged under a Peccei-Quinn symmetry that prevents any other supersymmetric interactions of $S$.  For weak interactions to break, the soft mass-squared parameter for $S,H_u$ or $H_d$ must scale negative, generating a new mass scale, $\mu_c$, by dimensional transmutation.  Taking $\mu_c \sim v$ would provide a natural understanding of the weak scale but, for $\xi\sim 1$, leads to a PQ breaking scale $f \sim v$.  This is experimentally excluded so this simple supersymmetric theory has not been studied before, except for the situation that $\xi \sim v/f \ll 1$ \cite{Feldstein:2004xi}.  With $\xi$ of order unity, $f \sim v$ is likely anthropically excluded because axion emission prevents main sequence stars having long lifetimes~\cite{Frieman:1987ui}.  

Hence, in Section \ref{sec:dimtrans} we study radiative PQ breaking via a dimensional transmutation induced by $m_S^2$ passing through zero at a scale $\mu_c$, giving 
\be 
\vev{S} \sim f \sim \mu_c \gg v
\ee
and take $10^{10} \GeV <  \;  \mu_c \; \ll M_*$.  Radiative PQ breaking requires $\tilde{m} \lsim \mu_c$, while electroweak symmetry breaking requires $\tilde{m} \gsim \mu = \xi \vev{S} \sim \mu_c$ so that $\tilde{m}$, which is taken to scan in the multiverse,  is selected to be of order order $\mu_c$.  Thus the superpartner mass scale, the PQ breaking scale and the dimensional transmutation scale are all comparable
\be
\tilde{m} \sim f \sim \mu_c.
\label{eq:scales}
\ee
For an alternative scheme with supersymmetry broken at an intermediate scale, and the possibility of an axion with $f \sim \tilde{m}$, see \cite{Hall:2013eko, Hall:2014vga}.

What is the effective theory below the scale $\tilde{m} \sim f \sim \mu_c$?  The anthropic necessity of electroweak symmetry breaking requires a single Higgs doublet, $h$, to be have a mass parameter fine tuned to be much less than $\mu_c$.  Thus this effective theory is the SM augmented by the axion supermultiplet.  In Section \ref{sec:EWSB} we study the scalar potential of this effective theory. The saxion is of particular interest as it has a squared mass, $m_s^2$, that is one-loop suppressed compared to $\mu_c^2$, and a scalar trilinear coupling $A \xi \,s h^\dagger h$, with $A$ expected to be order $\tilde{m}$.  On integrating out the saxion, a negative contribution to the Higgs quartic coupling results, $-\lambda_-$, that is parametrically of order $- \xi^2 A^2/m_s^2$ giving a SM Higgs quartic coupling
\be
\lambda \, = \, \lambda_+ - \lambda_- \, = \, \left(\frac{g^2 + g^{\prime\,2}}{8} \cos^2 2\beta \, + \,  \frac{\xi^2}{4} \sin^2 2\beta \right) \, - \, \left(  \frac{\xi^2A^2}{2 \, m_s^2} \right)
\label{eq:SMquartic}
\ee 
where $\lambda_+$ contains the positive $D$ and $F$ term contributions, and all couplings are renormalized at $m_s$.  
Since $A$ is typically of order $\tilde{m}$ while $m_s$ arises only at loop level, one expects $\lambda_- \gg \lambda_+$, leading to an instability of the SM electroweak vacuum \cite{Linde:1979ny,Lindner:1985uk,Sher:1993mf, Holthausen:2011aa,EliasMiro:2011aa,Degrassi:2012ry,Buttazzo:2013uya}.  

\begin{figure}
\begin{center}
\includegraphics[height=3.54in]{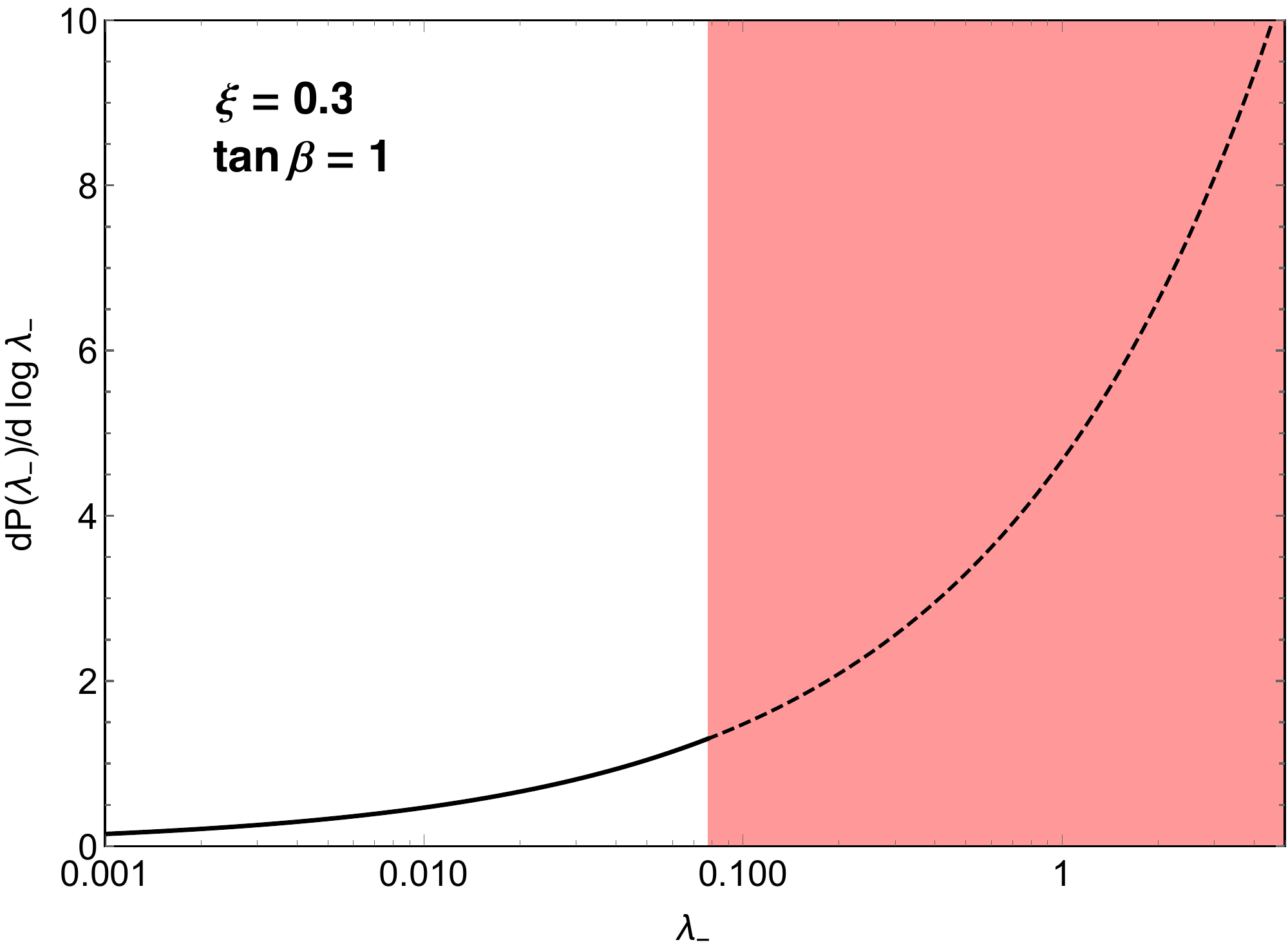}
\end{center}
\caption{The probability distribution for $\lambda_-$  at scale $m_s$, where the total Higgs quartic coupling is $\lambda = \lambda_+ - \lambda_-$.  The shaded region is excluded as the lifetime of the electroweak vacuum is less than $10^{10}$ years, with the boundary shown for $\xi = 0.3$ and $\tan\beta = 1$.}
\label{fig:probintro}
\end{figure}

Hence, in the multiverse most universes have no stable electroweak vacuum with $v \ll \mu_c$.   However, $A$ has two contributions that are typically comparable.  As the soft supersymmetry breaking parameters are scanned, there can be a cancellation in $A$ yielding a stable electroweak vacuum with $v \ll \mu_c$, which we study in Section \ref{sec:Higgspred}. The required cancellation in $A$ is mild, about 1 in 10, very much less than the cancellation required for a light Higgs doublet.  Assuming generic distributions for the relevant soft parameters, the distribution for $A$ in the region of the cancellation takes the form
\be
dP(A) \propto dA.
\label{eq:P(A)}
\ee  
This leads to a distribution for $\lambda_-$ at $m_s$ that is peaked towards large values, as shown in Figure \ref{fig:probintro}.  Since $\lambda_+$ is small and does not scan, this favors large negative values of the Higgs quartic $\lambda$ at $m_s$, explaining why typical observers lie close to the electroweak vacuum metastability boundary.  It is known that a multiverse distribution favoring low values of the quartic coupling at high scales leads to a prediction for the Higgs mass \cite{Feldstein:2006ce}; however the origin and strength of this distribution was unknown, leading to a large uncertainty in the prediction.  Nevertheless the physics was clear: the multiverse distribution for the Higgs quartic at high scales explains the close proximity of the Higgs boson mass to the bound that follows from requiring an electroweak vacuum with lifetime of order $10^{10}$ years.

In Section \ref{sec:Higgspred}, using the distribution shown in Figure \ref{fig:probintro} following from (\ref{eq:P(A)}), we compute a statistical prediction for the SM Higgs quartic coupling and therefore the Higgs boson mass.  This  prediction is shown as a function of $\xi$ by the black curves in Figure \ref{fig:ProbabilityLambda3}, for two values each of $\mu_c$ and $\tan \beta$.  The agreement with the observed Higgs mass, shown in red, is remarkable.  In the simple supersymmetric theory described by (\ref{eq:W}), for a wide range of $\xi$, $\mu_c$ and $\tan \beta$, the Higgs mass is correctly predicted within the $1 \sigma$ statistical error of about 5 GeV that originates from the multiverse distribution of Figure \ref{fig:probintro}.

While superpartner masses are of order $\tilde{m} \sim \mu_c$, the detailed spectrum is highly model-dependent.  However, the masses of the heavy Higgs doublet, the saxion and the axino are highly constrained, and are computed in Section \ref{sec:spect} as a function of $\xi$ and $\tan \beta$, in units of $\mu_c$.

In section 6 we argue that the theory yields single-component axion dark matter.  If $R$ parity is conserved, there is a strong upper bound on the reheat temperature after inflation to ensure that very large, and environmentally damaging, abundances of LSPs are not produced.  With rapid reheating after inflation we find that the PQ phase transition must occur before inflation, so that the dark matter abundance depends on the vacuum misalignment angle $\theta_i$.  The observed abundance, which fixes the dimensional transmutation scale to be $f \sim 10^{12}/\theta_i^2$ GeV, can be understood from a wide range of multiverse distributions for $\tilde{m}$.  For example, if this distribution is mild, the cost of fine-tuning the electroweak scale implies a preference for low values of $\tilde{m}$ and therefore $f$, so that the observed dark matter abundance is close to the minimal value that allows virialization of galactic halos \cite{D'Eramo:2014rna}.

In section 7 we show that the Higgs mass prediction of Figure \ref{fig:ProbabilityLambda3} applies to a large class of theories where the SM has a completion involving a SM gauge singlet scalar at scale $\Lambda > 10^{10}$ GeV.

A Higgs sector described by (1) is perhaps the simplest supersymmetric axion model.  One possibility for the PQ charges is: 1 on matter, -2 on $H_{u,d}$ and 4 on $S$. This assignment explains the absence of proton decay via operators of dimension 4 and 5, is consistent with $SU(5)$ and $SO(10)$ grand unification, and leads to $R$ parity conservation.  Neutrino masses could be Dirac, or the seesaw mechanism can be implemented from the vev of an addition field carrying both PQ and lepton charges.

\section{PQ Breaking from Dimensional Transmutation}
\label{sec:dimtrans}


The model is obtained by adding a SM gauge singlet chiral superfield $S$ to the MSSM field content. The interactions for $S$ are given by the superpotential of eq.~(\ref{eq:W}) together with the soft SUSY breaking potential
\be
V_{\textrm{soft}}\;=\; \xi A_\xi\, S H_u H_d \,+{\textrm{h.c.}}+\,m_S^2 \,|S|^2.
\ee
At tree level the resulting lagrangian has an exact $U(1)$ PQ global symmetry under which $S$ and $H_u H_d$ rotate with opposite charges, forbiding the appearance of a $\mu$-term and the corresponding soft SUSY breaking $B\mu$-term. 

We define the various parameters and fields at renormalization scale $M_*$, the cutoff of the theory, which we denote by a subscript.   Both $\xi_*$ and $A_{\xi*}$ can be taken to be positive with no loss of generality. According to the sign of the soft parameter $m_{S*}^2$ we can distinguish two cases. If $m_{S*}^2<0$ then the PQ preserving vacuum is unstable. The saxion vev is stabilized by higher dimensional operators and $\langle S\rangle\sim M_*$. We discard this possibility. If, on the other hand, $m_{S*}^2>0$ the PQ symmetry is unbroken at tree level\footnote{An order $\tilde{m}$ value of $A_\xi$ can drive PQ symmetry breaking. Notice that in this case one expects $\langle S\rangle\sim \langle H_u\rangle \langle H_d\rangle / \tilde{m}$ which is unacceptable.}. Once loop effects are included, however, the potential is modified for values of the field $S\ll M_*$ and a PQ breaking minimum can develop. 
This occurs because the superpotential coupling in eq.~(\ref{eq:W}) will generically drive the $S$ soft mass negative at a scale 
\be\label{muC}
\mu_c\sim M_* e^{-4\pi^2/\xi_*^2}
\ee
where gauge and top Yukawa couplings have been neglected and for illustration we take equal soft parameters for the scalars in $H_u, H_d$ and $S$ and for the trilinear interaction: $m^2_{H_u*} = m^2_{H_d*} =m_{S*}^2 = A^2_{\xi*}$.  The scale $\mu_c$ is a priori independent of the overall scale of supersymmetry breaking $\tilde{m}$ and is generated by dimensional transmutation. 

\begin{figure}
\begin{center}
\includegraphics[height=2.54in]{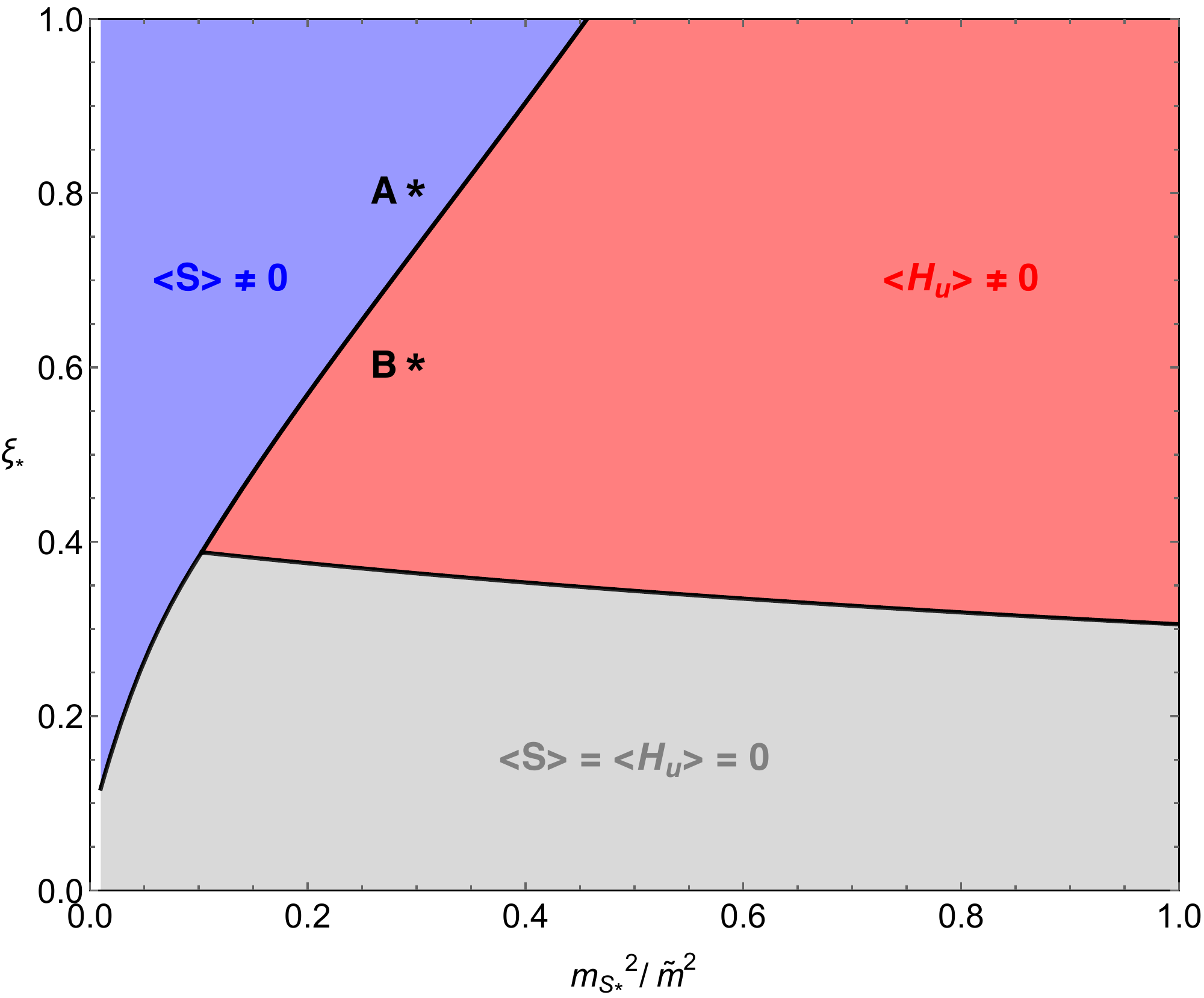} $\qquad$ \includegraphics[height=2.54in]{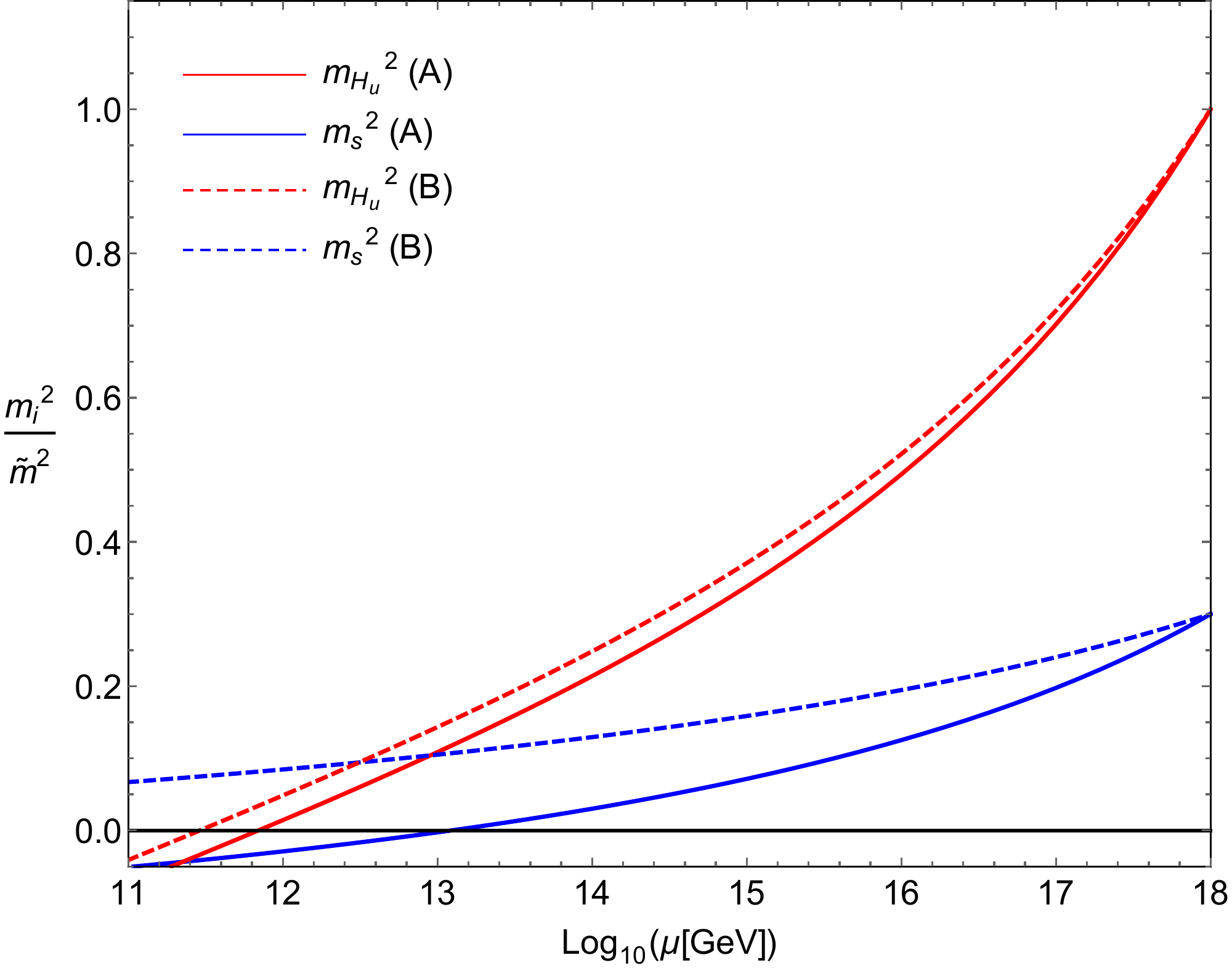}
\end{center}
\caption{Left panel: phase diagram of our model for $\tilde{m} = 10^{11}$ GeV, showing regions of PQ breaking (blue), EW symmetry breaking (red) and no symmetry breaking (gray). Right panel: running of the soft parameters $m_{H_u}^2$ and $m_S^2$ for the two benchmark points $A$ and $B$.}
\label{fig:PQbreaking}
\end{figure}

This radiative breaking of PQ symmetry is illustrated in Figure \ref{fig:PQbreaking}.  The left panel shows a phase diagram in the 
$(m^2_{S*}, \xi_*)$ plane for $\tilde{m} = 10^{11}$ GeV, with other relevant parameters fixed at $m_{H_u*} =  m_{H_u*} = m_{Q*} = m_{u*} = A_{\xi*} = A_{t*} = \tilde{m}$ and $y_{t*} =1$ for the top quark Yukawa coupling.  In the gray region neither electroweak nor PQ symmetry breaks, while in the red region electroweak symmetry breaks at a scale $\mu_c \gsim \tilde{m}$.  Hence it is the blue region, with PQ symmetry breaking at the dimensional transmutation scale  $\mu_c \gsim \tilde{m}$, that we study in this paper.  The right panel shows running of $m_{S}^2$ and $m_{H_u}^2$ for the benchmark points A and B defined in the right panel.

The vev of $S$ is computed by minimizing the full effective potential for $S$, which at 1-loop leading-log order is
\be\label{VLL}
V(S)=\Lambda(\mu)+m_S^2(\mu)|S(\mu)|^2+ V^{(1)}(S,\mu)
\ee
where $V^{(1)}$ is the 1-loop Coleman-Weinberg~\cite{Coleman:1973jx} potential
\be
V^{(1)}(S,\mu)=\sum_i (-1)^{2s_i} \frac{m_i^4(S)}{64\pi^2}\left(\log\frac{m_i^2(S)}{\mu^2}-\frac{3}{2}\right)
\ee
where the sum extends over all particles $i$ in the model with $m_i(S)$ being their field dependent masses and $s_i$ their spins. Given the leading-log accuracy at which we are working, all the various parameters appearing in $V^{(1)}$ should be evaluated at a reference scale $\mu_0$ such that $\log \mu/\mu_0$ is not much larger than 1. The running parameters in eq.~(\ref{VLL}) are evaluated at the running scale $\mu$ according to their RG equation. The potential $V$ is $\mu$ independent at the order at which we are working. This can be explicitly checked using
\be
16\pi^2\frac{d\Lambda}{d\log\mu}=2m_{H_u}^4+2m_{H_d}^4+m_S^4,~~~ 16\pi^2 \frac{d \log S}{d\log\mu}=-2\xi^2.
\ee
The running cosmological constant is included to cancel the $\mu$ dependence of the field independent part of $V$. 

It is extremely convenient to choose the scale $\mu_0$ to be the renormalization point $\mu_c$ at which $m_S^2(\mu_c)=0$, so that the leading-log effective potential becomes
\be\label{VS}
V(S)=\frac{1}{64\pi^2}\left[4 m_H^4 \left(\log\frac{m_H^2}{\mu_c^2}-\frac{3}{2}\right)+4 m_h^4\left( \log\frac{m_h^2}{\mu_c^2}-\frac{3}{2}\right)-8 m_F^4\left( \log\frac{m_F^2}{\mu_c^2}-\frac{3}{2}\right)\right]
\ee
where the parameters $m_H^2$, $m_h^2$ and $m_F$ depend on $S$ through
\bea
m_{H,h}^2&=&\frac{m_{H_u}^2+m_{H_d}^2}{2}+\xi^2 |S|^2\pm\sqrt{\left(\frac{m_{H_u}^2-m_{H_d}^2}{2}\right)^2+\xi^2 A_\xi^2 |S|^2}\\
m_F&=&\xi |S|.
\eea

All soft masses and couplings are now evaluated at $\mu_c$ so that the relevant parameter set is 
\be\label{param}
\mu_c,~~\xi(\mu_c),~~m_{H_u}^2(\mu_c),~~ m_{H_d}^2(\mu_c),~~ A_\xi(\mu_c).
\ee
The scale $\mu_c$ arises from dimensional transmutation, the soft parameters $m_{H_u}, m_{H_d}, A_\xi$ are of order the supersymmetry breaking scale $\tilde{m}$, and we take $\xi$ to be order unity.  Minimizing $V(S)$ of (\ref{VS}) leads to $\xi\langle S\rangle\sim \mu_c$, so that the axion decay constant is given by the dimensional transmutation scale $f = \langle S\rangle\sim \mu_c$. This generates a supersymmetric Higgs mass parameter $\mu = \xi\langle S\rangle\sim \mu_c$ which is unrelated to the scale of supersymmetry breaking.  If $\mu_c \gg \tilde{m}$ electroweak symmetry is unbroken;  if $\mu_c \ll \tilde{m}$ radiative PQ breaking fails, as the superpartners become massive before $m_S^2$ scales negative, so that either electroweak symmetry is unbroken, or PQ symmetry breaks at the weak scale.  Hence, the viability of this model requires a special choice $\tilde{m} \sim \mu_c$; furthermore, since $f \sim \mu_c$ supersymmetry breaking occurs many orders of magnitude above the weak scale, which is therefore highly fine-tuned.   In the next section we take the soft supersymmetry breaking parameters at $M_*$ to scan in the multiverse and study the consequences of imposing an environmental requirement of electroweak symmetry breaking.  The entire relevant parameter set of (\ref{param}) scans; however, since $\xi(M_*)$ does not scan, $\xi(\mu_c)$ scans only via its $\mu_c$ dependence which is logarithmic and mild.

\section{Higgs Quartic Coupling of SM}
\label{sec:EWSB}
To obtain the SM Higgs potential we must identify the SM Higgs state and obtain the effective theory by integrating out super partners and heavy scalars.

In a supersymmetric theory in which there is a hierarchy between the scale of SUSY breaking and the weak scale $\tilde m \gg m_Z$, the condition for EWSB can be expressed as $\det \mathcal M_H^2\approx 0$, where $\mathcal M_H^2$ is the Higgs mass matrix around the EW preserving (but PQ breaking) vacuum
\be
\mathcal M^2_H =\begin{pmatrix}
\mu^2+ m_{H_u}^2 & A_\xi\,\mu \\
A_\xi\, \mu&\mu^2+ m_{H_d}^2
\end{pmatrix}
\ee
where we defined $\mu\equiv \xi \langle S\rangle$.\footnote{We assume Im\,$\langle S\rangle=0$ without loss of generality.} Furthermore, with $\xi$ of order unity $\mu \sim \mu_c$.
As discussed above, the requirements of both PQ and EWSB relate the two a priori unrelated scales $\mu_c$ and $\tilde m$.

On fine tuning $\det \mathcal M_H^2\approx 0$, it is convenient to rotate from the $(H_u,\, H_d)$ basis to the $(h,\, H)$ one, where $h$ is the massless SM Higgs doublet:
\bea
H_u&=&\sin\beta h - \cos\beta H\,,\\
H_d&=&\cos\beta \tilde h + \sin\beta \tilde H\,,\\
\tan\beta&=&\sqrt{\frac{\mu^2+m_{H_d}^2}{\mu^2+m_{H_u}^2}}\,.
\eea
We defined $\tilde h=i\sigma_2 h^*$ and similarly for $ \tilde H$. 

\begin{figure}
\begin{center}
\includegraphics[scale=0.48]{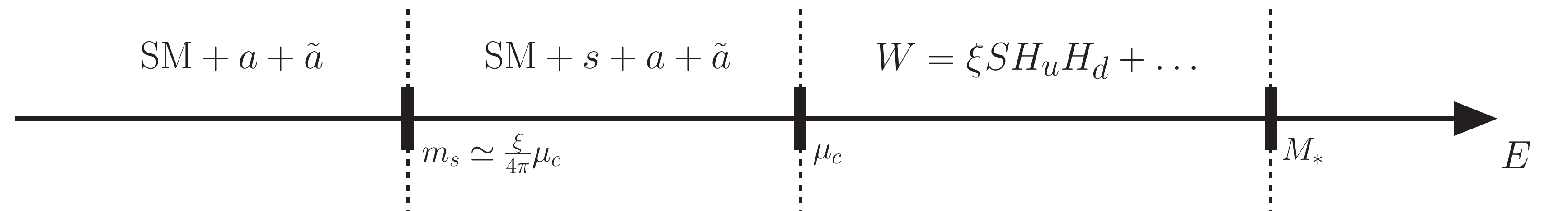}
\end{center}
\caption{The Effective Field Theories below $M_*$, $\mu_c$ and the saxion mass $m_s$. }
\label{fig:MassThresholds}
\end{figure}

As shown in Figure \ref{fig:MassThresholds}, on decoupling the superpartners and expanding around the $H=0$, $S=v_S$ vacuum, the effective theory below $\mu_c$ is the SM augmented by the axion and saxion fields (and an axino which we discuss later).   The scalar potential for this effective theory is
\be\label{Vhs}
V(h,s)=\frac{m_s^2}{2} s^2+A \xi \, s h^\dagger h+ \lambda_s \, s^4+ \lambda_{sh} \, s^2 h^\dagger h + \lambda_h \, (h^\dagger h)^2.
\ee
The mass of the saxion $s$ is obtained by expanding the Coleman-Weinberg potential of eq.~(\ref{VS}) around its minimum, leading to a value which is loop-suppressed compared to the scale of $\tilde{m}$ and $\mu$
\be
m_s^2\sim \frac{\xi^2}{(4\pi)^2} \, \mu^2
\label{eq:saxion mass}.
\ee
Furthermore there is a trilinear scalar interaction between $s$ and $h$, $A\xi \, sh^\dagger h$, where
\be\label{cubic}
A= \sqrt{2} \left( \mu - \frac{A_\xi}{2} \sin2\beta \right).
\ee
The various quartic couplings are fixed by supersymmetry and are positive sums of $D$- and $F$- term contributions.  Matching at tree-level at the scale $\mu_c$ gives
\be\label{quartich}
\lambda_h=\frac{g^2 + g^{\prime\,2}}{8} \cos^2 2\beta + \frac{\xi^2}{4} \sin^2 2\beta 
\ee
with all couplings evaluated at the scale $\mu_c$.

As shown in Figure \ref{fig:MassThresholds}, on integrating out the saxion, the effective theory below $m_s$ is the SM augmented by the axion with scalar potential 
\be
V(h)=\lambda (h^\dagger h)^2. 
\ee
Matching the theories at tree level at the boundary at $m_s$ gives
\be\label{quarticSM}
\lambda(m_s) = \lambda_h(m_s) -  \frac{(A\xi)^2}{2 \, m_s^2}
\ee
where the last term arises from a diagram with a virtual $s$ and two trilinear interactions and we ignore the running of $A\xi$ and $m_s$ between $\mu_c$ and $m_s$.
The important point is that this term is negative-definite and very large due to the lightness of $s$.  The natural expectation is that $(A \xi)^2/m_s^2 \sim 16 \pi^2$, so that $\lambda(m_s) \ll 0$ and there is no electroweak vacuum with $\langle h\rangle \ll m_s$.  

However, we insist on an environmental requirement of electroweak symmetry breaking at a scale close to that observed.  To accomplish this we scan the soft supersymmetry breaking parameters at $M_*$.  This allowed us to take one combination of the Higgs doublets much lighter than $\mu_c$, but this tuning alone is insufficient.  If no universes have the required electroweak symmetry breaking then our theory is excluded.  However, some universes do have the desired weak scale, those that have a cancellation between the two terms in (\ref{cubic}) so that $(A\xi)^2/m_s^2 \sim 1$ (recall that $\mu \sim \mu_c$, which scans with the soft parameters)\footnote{$A \simlt m_s$ also avoids having a deeper minimum of the potential (\ref{Vhs}) at $\langle s\rangle \sim \langle h\rangle \sim A/\xi$.}.  This is only a 1 in 10 fine-tune, much milder than that required to get a light Higgs doublet.  This fine-tuning, like that for $m_h$, is an inevitable consequence of our theory.   The necessity of this fine-tuning has an important consequence: it allows us to predict the probability distribution for the Higgs quartic coupling and therefore the Higgs boson mass, as we discuss next.

\section{Predicting The Higgs Quartic from Vacuum Stability}
\label{sec:Higgspred}

For generic values of the various soft parameters entering the Higgs sector, the SM Higgs quartic in eq.~(\ref{quarticSM}) is negative and much too large to lead to an acceptable electroweak vacuum state: for order one $\tan\beta$, the negative contribution coming from integrating out the saxion field is between one and two orders of magnitude bigger than the vacuum stability bound \cite{Linde:1979ny,Lindner:1985uk,Sher:1993mf}.

Let us rewrite the quartic boundary condition at $m_s$ of eq.~(\ref{quarticSM}) as
\be
\lambda(m_s) = \lambda_+ - \lambda_-,  \hspace{0.5in}  \lambda_- = \lambda_0 \epsilon^2.
\ee
$\lambda_+= \lambda_h(m_s)$ is obtained by scaling (\ref{quartich}) from $\mu_c$ to $m_s$. Including the leading scaling from the top quark coupling gives
\be\label{lambda+}
\lambda_+= \left( \frac{g^2 + g^{\prime\,2}}{8} \cos^2 2\beta + \frac{\xi^2}{4} \sin^2 2\beta 
+ \frac{3 y_t^4}{8 \pi^2} \ln(4 \pi/\xi) \right)_{\mu_c}.
\ee
The dimensionless quantity 
\be\label{epsilon}
\epsilon\equiv \frac{A}{\mu} = \sqrt{2} \left( 1 - \frac{A_\xi}{2 \mu} \sin2\beta \right)_{\mu_c}
\ee
 and $\lambda_0$ is a large coupling of order $(4\pi)^2$, which is defined from eq.~(\ref{quarticSM}). 

The soft supersymmetry breaking parameters are taken to scan at the cutoff $M_*$.  This leads to only a very restricted scanning of $\lambda_+$. There is a very mild logarithmic scanning of $(g,g',\xi, y_t)$ through the scale $\mu_c$ which scans.  As $\tan \beta$ scans, it induces only mild scanning of $\lambda_+$ as $g,g',\xi$ are roughly comparable; also, $\lambda_+$ has an upper bound because $\cos^2\beta + \sin^2 \beta =1$.  

On the other hand, $\lambda_-$ and $\epsilon$ scan strongly as they depend directly on the soft parameters.  While $\epsilon$ is naturally of order one and can have both signs, its anthropically allowed range is between one and two orders of magnitude smaller, depending on $\xi$, so that an environmental cancellation is forced on the right-hand side of (\ref{epsilon}). Due to this accurate cancellation, assuming generic pdfs for the various soft parameters, the probability distribution for $\epsilon$ in the anthropically relevant range can be accurately approximated by a flat prior,\footnote{This same argument is used in the literature to obtain the a priori distribution for the cosmological constant which is then used to discuss its anthropic constraints.}
\be\label{pdfepsilon}
dP(\epsilon)\propto d\epsilon.
\ee
The corresponding prior distribution for $\lambda_-$ is
\be
dP(\lambda_-) \, \propto \, \lambda_-^{-1/2} \; d\lambda_-\,.
\label{eq:priorlambda_-}
\ee
Since the scanning of $\lambda_+$ is so mild, this allows us to calculate the probability distribution for observed values of $\lambda(m_s)$.

The requirement that the SM electroweak vacuum has a lifetime longer than $10^{10}$ years, against quantum tunneling at scale $m_s$, leads to a constraint $\lambda(m_s)>\lambda_{\textrm{cr}}(m_s)$, and $\lambda_{\textrm{cr}}$ is known accurately \cite{Holthausen:2011aa,EliasMiro:2011aa,Degrassi:2012ry,Buttazzo:2013uya}.  In our theory this is interpreted as an anthropic constraint on $\lambda_-$
\be 
\lambda_-<\lambda_+-\lambda_{\textrm{cr}}.
\label{eq:constonl-}
\ee
In this region that allows observers the normalized distribution is
\be
dP(\lambda_-) \, = \, \frac{1}{2} \, \frac{\lambda_-^{-1/2}}{\sqrt{\lambda_+ - \lambda_{\textrm{cr}}}} \; d\lambda_-\,.
\label{eq:normpriorlambda_-}
\ee

With $d P(\lambda_-)$ in hand, we compute the average value of the Higgs quartic and its variance 
\be\label{averagelambda}
\vev{\lambda} =\frac{2}{3} \lambda_+  + \frac{1}{3} \lambda_{\rm cr},\qquad
\sigma_\lambda =\frac{2}{3\sqrt 5} \left|\lambda_+ - \lambda_{\rm cr} \right| 
\ee
where $\sigma_\lambda^2=\vev{\lambda^2} - \vev{\lambda}^2$. Eq.~(\ref{averagelambda}) shows that the average value of $\lambda$ gets farther and farther from its catastrophic value $\lambda_{\textrm{cr}}$ and asymptote to $\langle \lambda\rangle\sim \tfrac{2}{3}\lambda_+$.

\begin{figure}
\begin{center}
\includegraphics[scale=0.405]{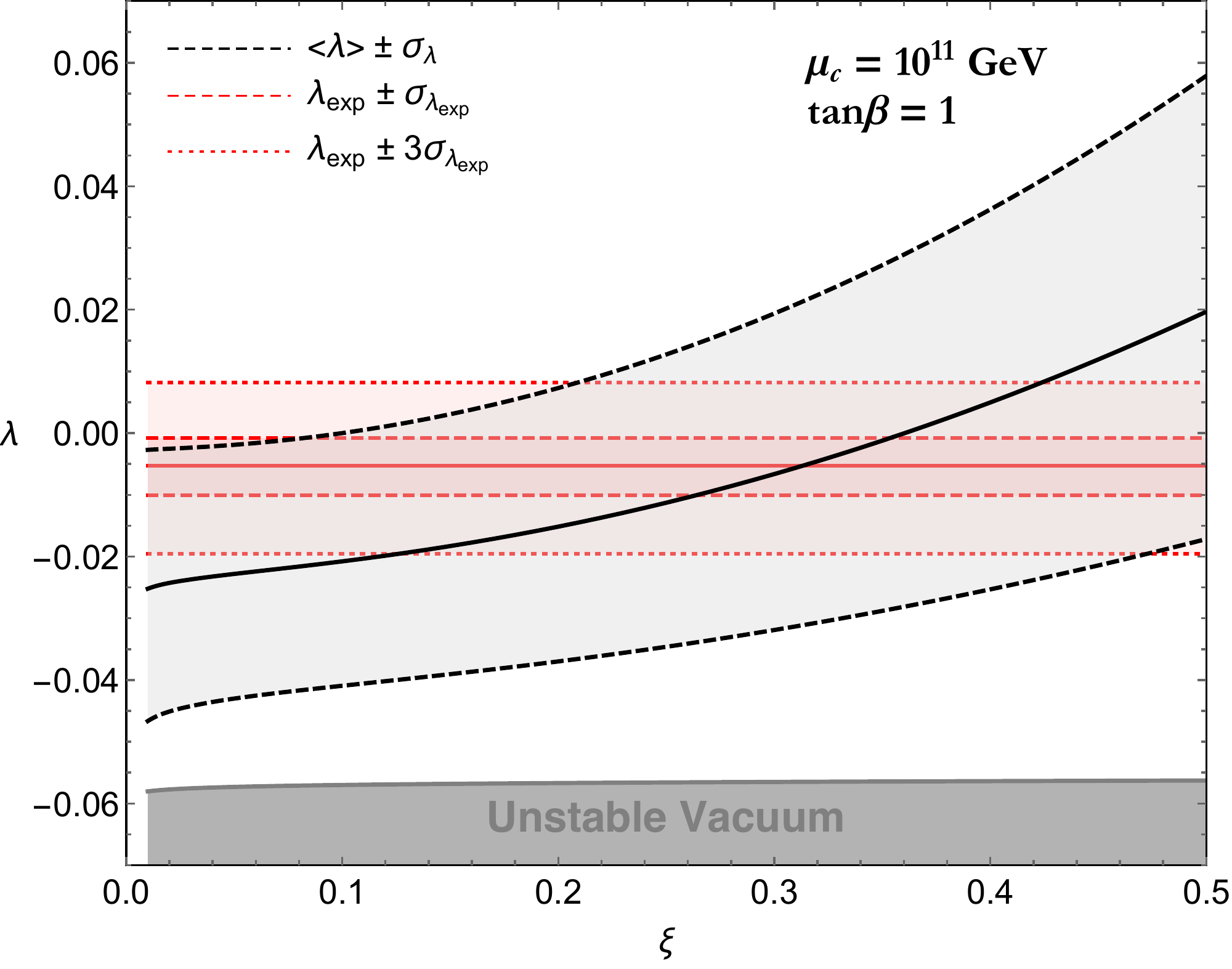} $\quad$ \includegraphics[scale=0.405]{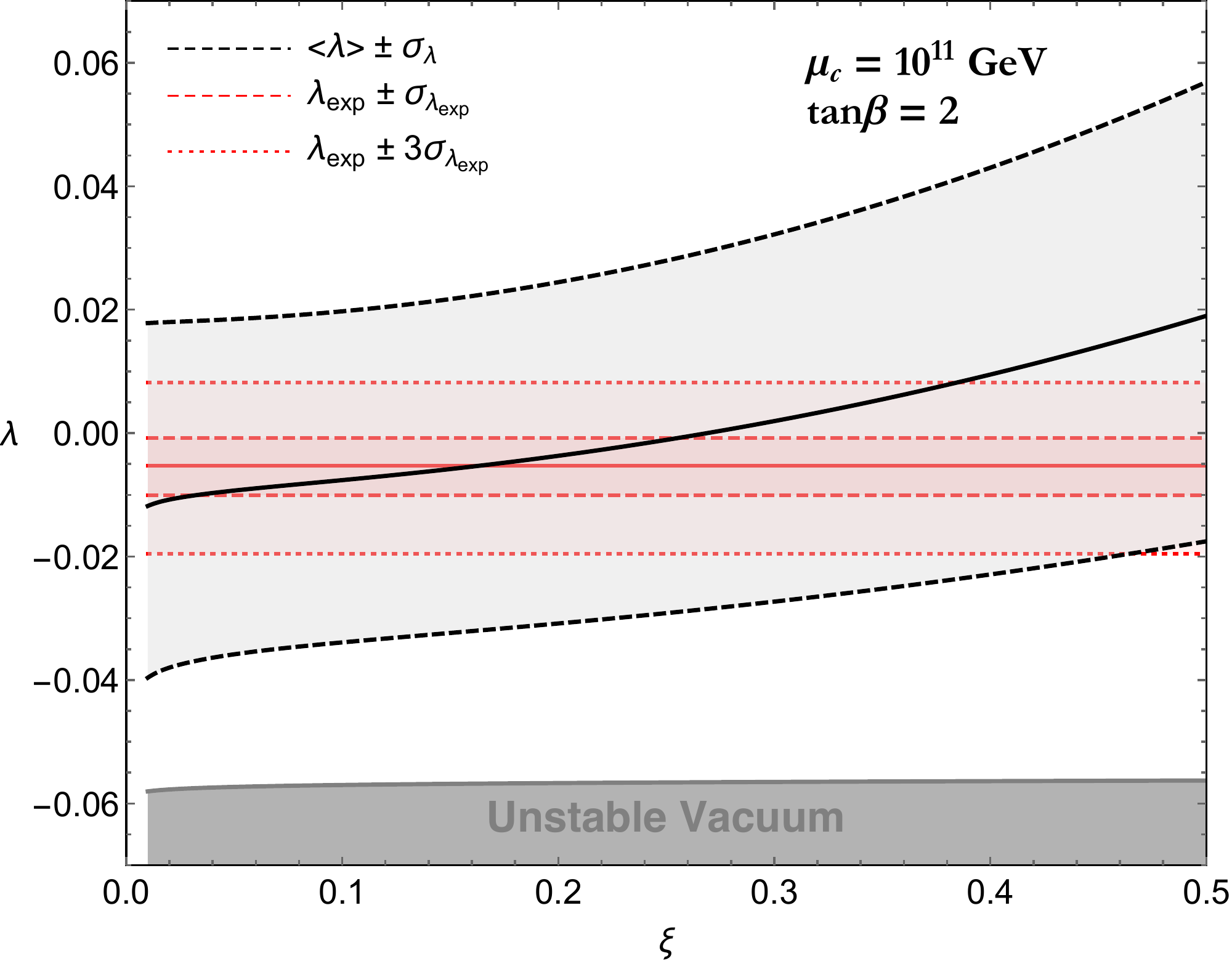}
 \\ \vspace{0.6cm}
 \includegraphics[scale=0.405]{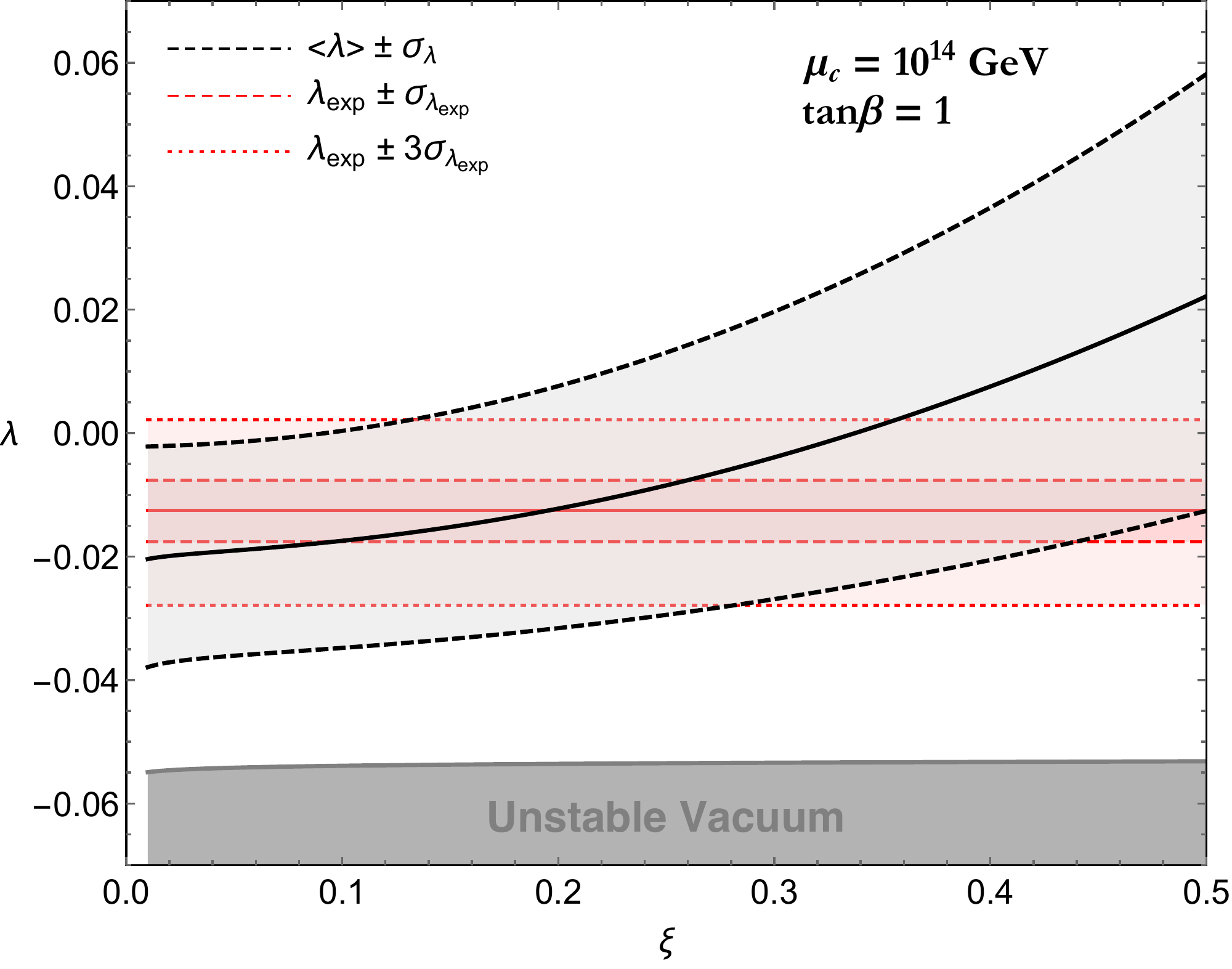} $\quad$ \includegraphics[scale=0.405]{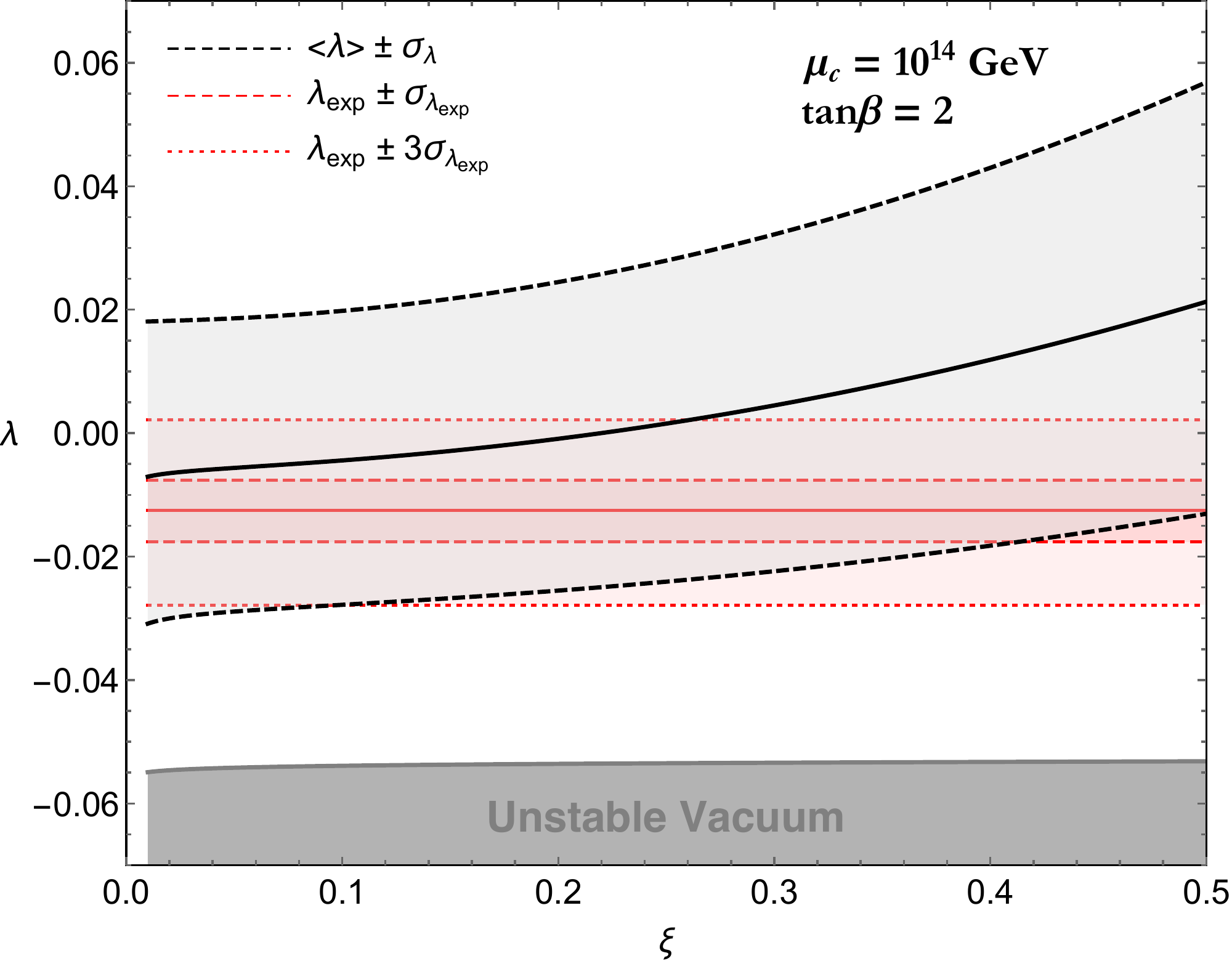}
\end{center}
\caption{Statistical prediction for the SM Higgs quartic coupling $\lambda(\mu_c)$ as a function of $\xi(\mu_c)$ for $\mu_c = 10^{11} \GeV$  (first row) and $\mu_c = 10^{14} \GeV$  (second row), and for $\tan\beta=1$ (left column) and $\tan\beta=2$ (right column). The solid line gives the average predicted value, while the dashed lines show the $1 \sigma$ statistical uncertainties in the prediction. The red shaded region is allowed by experiment at $1\sigma$ and $3\sigma$.}
\label{fig:ProbabilityLambda3}
\end{figure}

In Fig.~\ref{fig:ProbabilityLambda3} we plot the mean and $1\sigma$ range of $\lambda(\mu_c)$ as a function of $\xi$.  The values of eq.~(\ref{averagelambda}) are scaled up from $m_s$ to $\mu_c$ using SM RG equations so that all quantities in Fig.~\ref{fig:ProbabilityLambda3} refer to fixed values of $\mu_c = 10^{11} \GeV \; (10^{14} \GeV) $ in the left (right) panel. We include the 1-loop top  contribution from (\ref{lambda+}), which is visible in the prediction at low $\xi$.  The theoretical extrapolation to $\mu_c$ of the low-energy Higgs quartic coupling obtained from the Higgs mass measurement~\cite{Buttazzo:2013uya} is shown by the red band.  The agreement of the Higgs mass prediction (black) with data (red) is striking. In particular, very small values of $\lambda(\mu_c)$ in the range from $-0.02$ to $+0.01$ can be understood from values of $\xi_*$ that are order unity. The size of our $1\sigma$ prediction on the Higgs boson mass can be estimated from the the uncertainty on the Higgs quartic running due to the experimental uncertainty on the measured Higgs boson mass~\cite{Buttazzo:2013uya}. We find that the $1\sigma$ band at the scale $\mu_c$ translates into an uncertainty on the Higgs mass prediction of the order of $5 \GeV$.

\section{The spectrum of the model}
\label{sec:spect}

\begin{figure}
\begin{center}
\includegraphics[scale=.57]{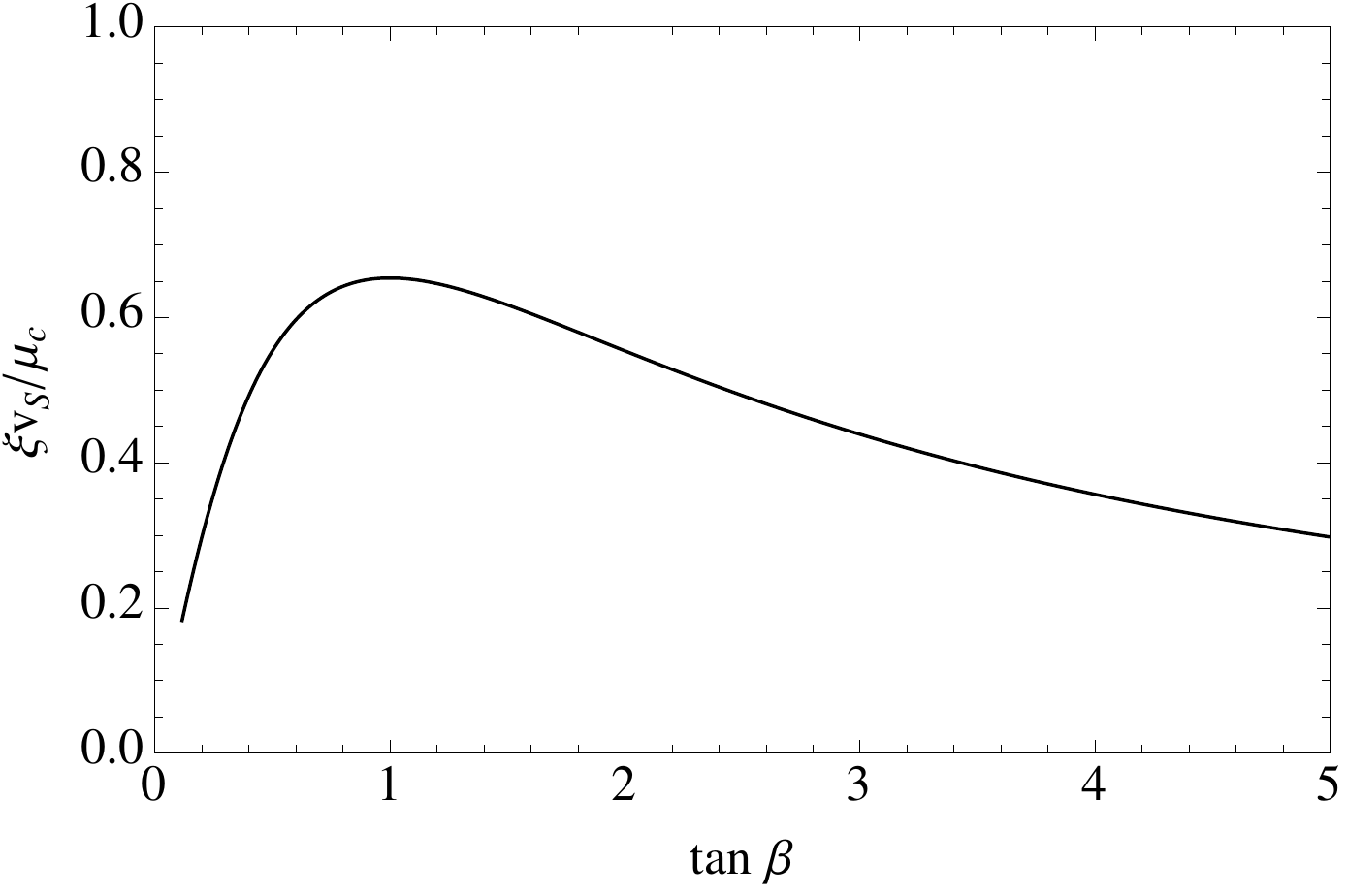}~~~\includegraphics[scale=.57]{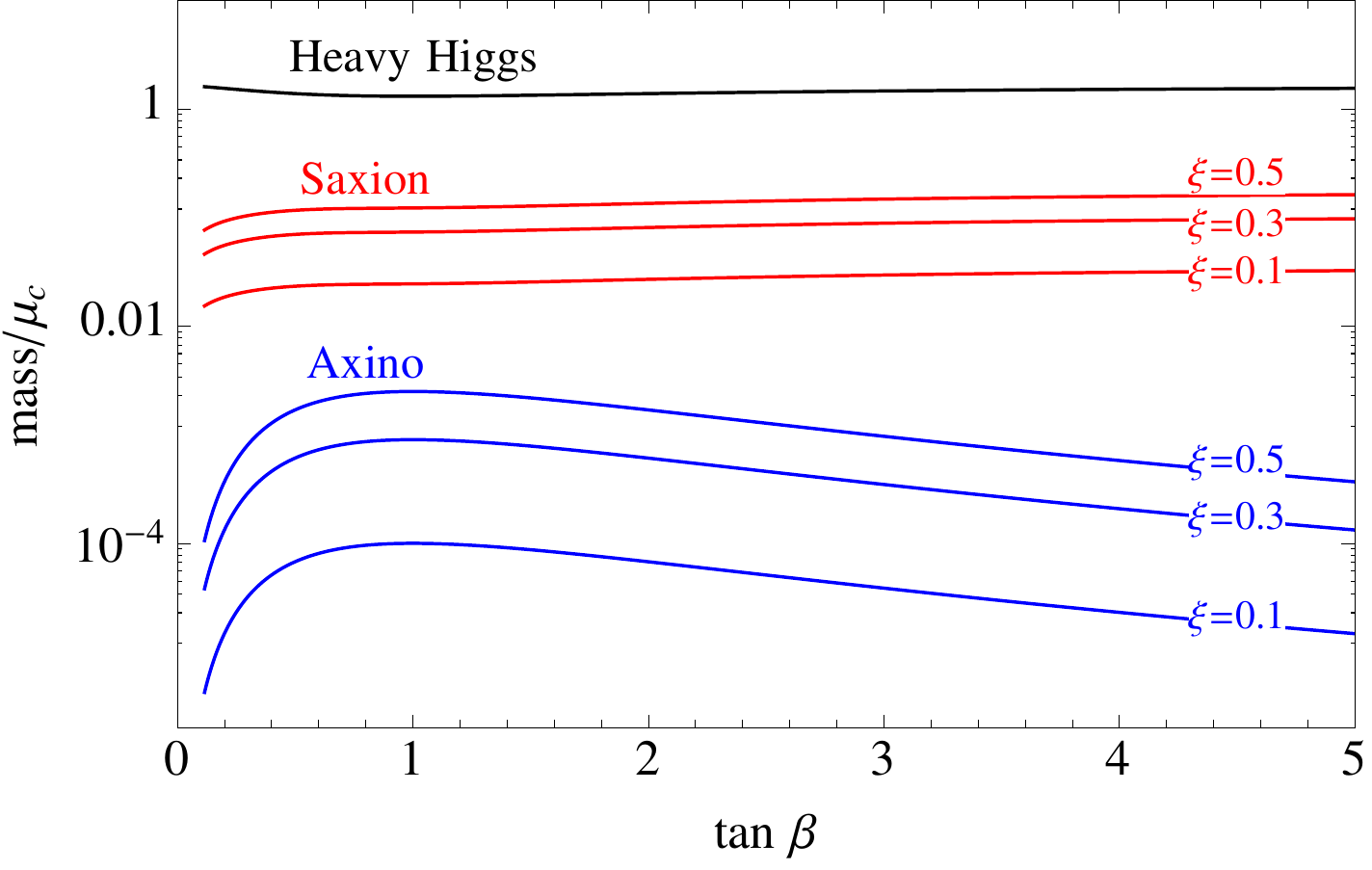}
\end{center}
\caption{Left panel: ratio $\xi v_S/\mu_c$ as a function of $\tan\beta$. Right panel: heavy Higgs, saxion and axino masses normalized to the scale $\mu_c$ as a function of $\tan\beta$. The saxion and axino masses are also shown for three reference values of $\xi$.}
\label{fig:xiS}
\end{figure}

The three input parameters $m_1^2$, $m_2^2$ and $A_\xi$ can be traded for the more physical ones $\det \mathcal M_H^2$, $A$ and $\tan\beta$. $\det \mathcal M_H^2$ and $A$ are furthermore constrained by the anthropic requirements coming from EWSB. To simplify the discussion of the model we will fix both of them to be zero, even though this is not strictly necessary for $A$. Having done this the Higgs sector of the model is completely defined by three parameters $\mu_c$, $\xi(\mu_c)$ and $\tan\beta$. In the left panel of Fig.~\ref{fig:xiS} we plot the ratio $\mu/\mu_c\equiv \xi v_S/\mu_c$ as a function of $\tan\beta$. As anticipated in the previous section this ratio is $O(1)$ for all relevant values of $\tan\beta$. Furthermore it is independent of $\xi$ as the parameter enters the lagrangian only through the combination $\xi S$.

In the right panel of Fig.~\ref{fig:xiS} we show, as a function of $\tan\beta$, the value of the heavy Higgs, saxion and axino masses, normalized to the scale $\mu_c$. The axino mass which is generated at tree level is tiny
\be
m^{\textrm{tree}}_{\tilde a}=\frac{\xi^2 v^2}{\mu}\sin2\beta
\label{axinotree}
\ee
coming from integrating out the higgsino. The axino mass receives a much larger one-loop contribution
\be
m^{\textrm{1-loop}}_{\tilde a}=\frac{\xi^2}{8\pi^2}\mu\sin 2\beta \frac{m_A^2}{\mu^2-m_A^2}\ln \frac{|\mu|^2}{m_A^2}.
\label{axinoloop}
\ee
This loop is analogous to the one-loop higgsino threshold contribution to the bino and wino masses in anomaly mediation~\cite{Giudice:1998xp,Gherghetta:1999sw}.  The axino typically receives a contribution of order the gravitino mass from higher dimensional operators \cite{Cheung:2011mg} which we do not include here as the gravitino mass is model-dependent.  

\begin{figure}
\begin{center}
\includegraphics[scale=.75]{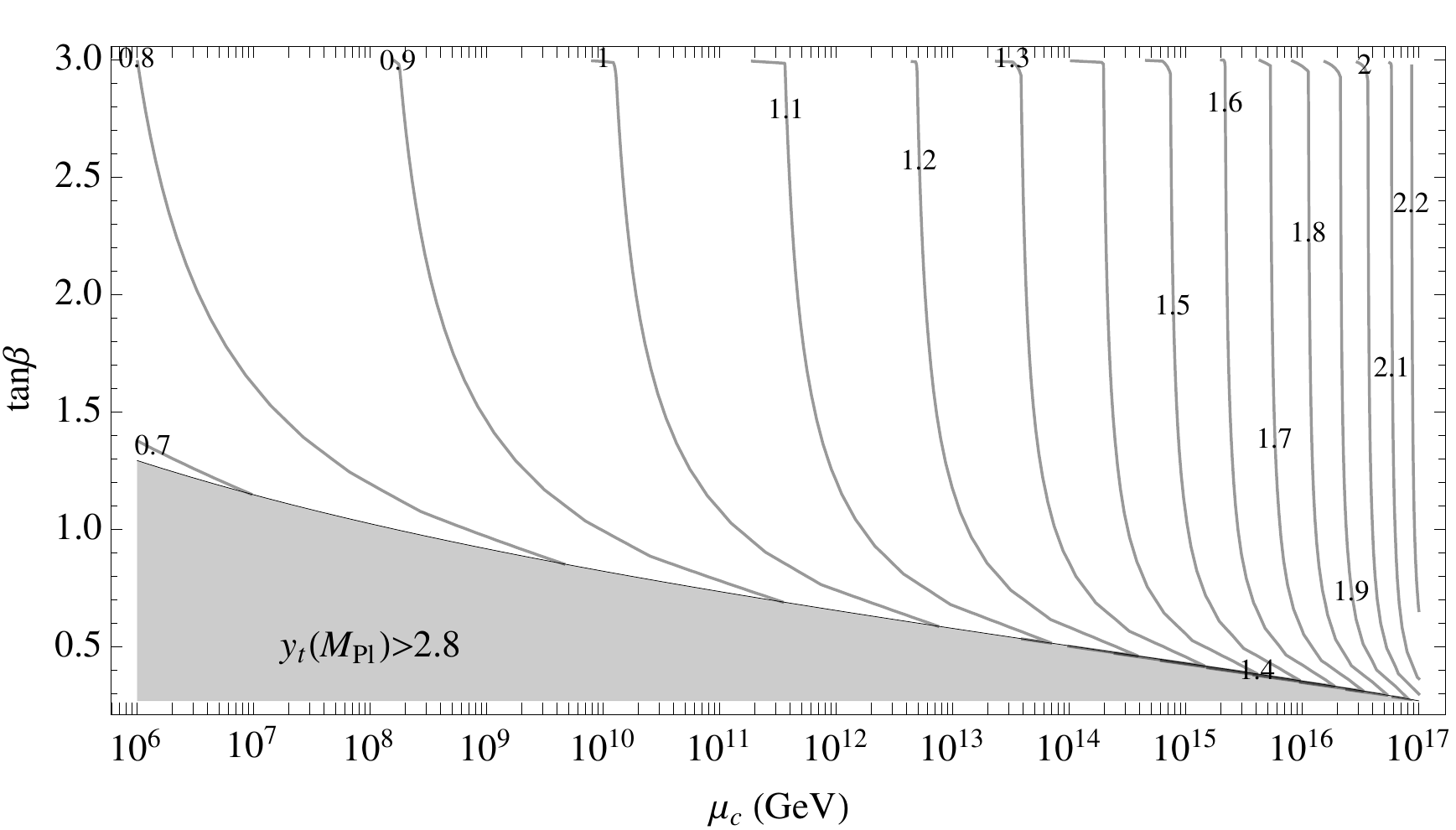}
\end{center}
\caption{Maximal value of $\xi$ at the scale $\mu_c$ under the assumption of a maximal allowed value of $\xi$ at $M_{Pl}$. In the gray region $3y_t^2/8\pi^2>0.3$ at the Planck scale.}
\label{perturbativity}
\end{figure}

Since the RG evolution makes $\xi$ grow in the UV, the allowed values of $\xi$ can be constrained from above by the requiring it to be in the perturbative range up to the cutoff scale $M_*$. In Fig.~\ref{perturbativity} we show the maximal value that $\xi$ can attain at the scale $\mu_c$ imposing that $\xi^2/4\pi^2<0.3$ at the Planck scale. This maximal value depends on $\tan\beta$. In Fig.~\ref{perturbativity} we also require the top Yukawa coupling to be perturbative up to the Planck scale ($3 y_t^2/8\pi^2<0.3$), which eliminates the small $\tan\beta$ and small $\mu_c$ region.

Arbitrarily small values are in principle allowed for $\xi$. Notice however that very small values of $\xi$, other than being difficult to explain from a theoretical point of view, induce a very slow running of $m_S^2$ from the cutoff scale $M_*$ down to the dimensional transmutation scale $\mu_c$. This implies that the UV spectrum of theories with very small $\xi$ is rather peculiar, with $m_S^2$ being much smaller than the other superpartners soft masses.
\newline

As the PQ symmetry of the model is broken by the vev of $S$ the theory contains a massless (at tree-level) boson. We want to identify this Goldstone boson with the QCD axion. In order to study its interaction it is useful to parametrize the neutral components of $S$, $H_u$ and $H_d$ as
\be
 S=\left(v_{S}+\frac{s}{\sqrt 2}\right)e^{i\frac{\phi_{S}}{\sqrt{2}v_{s}}},\quad H^0_{i}=\left(v_{i}+\frac{h_{i}}{\sqrt 2}\right)e^{i\frac{\phi_{i}}{\sqrt{2}v_{i}}},\quad i=u,d.
\ee
The phases $\phi_{s,1,2}$ are a valid parametrization for the pseudoscalar excitations of the theory and $v_S\equiv \langle S\rangle$. Only one linear superposition of the $\phi$'s gets a tree level mass in the theory. The relevant combination is easily identified as the only dependence on the phases of the various field in the potential is through the $A$-term for $\xi$
\be
A_\xi S H_u^0 H_d^0+{\textrm h.c.}\supset 2 A_\xi v_u v_d v_S \cos\left(\frac{\phi_u}{\sqrt2v_u}+\frac{\phi_d}{\sqrt2v_d}+\frac{\phi_S}{\sqrt2v_S}\right).
\ee
We can thus write
\bea
G&\equiv&\cos\beta \phi_d -\sin\beta \phi_u\\
A&\equiv&(\sin\beta \phi_d+\cos\beta\phi_u)\cos\alpha-\sin\alpha\phi_S\\
a&\equiv&(\sin\beta \phi_d+\cos\beta\phi_u)\sin\alpha+\cos\alpha\phi_S
\eea
where $G$ is the linear combination eaten by the $Z$ boson, $A$ is the heavy pseudoscalar and $a$ is the massless axion. We also have
\be
\tan\beta=\frac{v_u}{v_d}~~~~~{\textrm{and}}~~~~~\tan\alpha=-\sin\beta\cos\beta\frac{v}{v_S}.
\ee 
Under the following rephasing of the fields
\bea
H_{i}\to e^{iQ_i\alpha} H_i,~ S\to e^{iQ_s\alpha} S ~~~{\textrm{with}}~~~ Q_1=-\sin^2\beta Q_S,~ Q_2=-\cos^2\beta Q_S
\eea
one has
\be\label{shift}
G\to G,~~ A\to A,~~ a\to a+f Q_S\alpha
\ee
where we defined
\be
f=\sqrt2\,v_S\cos\alpha-\sqrt 2\,v\sin\alpha\sin\beta\cos\beta=\sqrt 2\frac{v_S}{\cos\alpha}\approx \sqrt 2 v_S.
\ee
In the following we fix $Q_S=1$. The field $a$ is thus the canonical axion field, shifting in a simple way under a PQ transformation. We refer to Appendix~\ref{axioninteractions} for a discussion of the leading interactions of the axion field in our model.


\section{Three Scenarios for Axion Dark Matter}
\subsection{Dark Matter Overview}
\label{subsection:DMO}

In general we might expect dark matter to have both axion and LSP components.  In this section we argue that dark matter is entirely axionic, and results from three distinct multiverse scenarios with differing values of $f$.

The superpartner mass parameters at the cutoff scale $M_*$ scan in our theory. It is convenient to define $\tilde{m}$ as the typical soft mass relevant for the fine tuning of the weak scale and, after marginalizing over the condition that a sufficiently stable electroweak vacuum exists, we assume a power-law distribution for $\tilde{m}$. The dynamically generated scale $\mu_c \sim \xi \langle S \rangle$ also scans in the multiverse, and in principle is unrelated to the SUSY breaking scale. However, as extensively discussed in Sec.~\ref{sec:EWSB}, the anthropic requirement of successful EWSB imposes $\mu_c \sim \tilde{m}$. In this Section we investigate the scanning of $\tilde{m}$ in the multiverse and its consequence for dark matter.

Although $\xi$ does not scan it is a free parameter and, furthermore, there are other parameters arising from inflation that affect the dark matter abundance.  An important distinction is whether the (last) PQ phase transition occurred before or after inflation, leading to two different cosmological axion scenarios: ``Pre-Inflation" and ``Post-Inflation". The axion is not present during inflation if the PQ breaking scale $f$ is lower than the Gibbons-Hawking temperature
\be
f \leq T_{\rm GH} \equiv \frac{H_I}{2 \pi} \ , \qquad \qquad \qquad \qquad H_I = \sqrt{\frac{8 \pi}{3}}  \frac{E_I^2}{M_{\rm Pl}} \ ,
\ee
with $E_I$ the energy scale of inflation. Even if this is not the case, PQ symmetry may be restored afterwards if $f < T_{\rm max}$, with $T_{\rm max}$ the maximum temperature attained during the reheating process. Thus we define Pre-inflation cosmology from the requirement
\be
f \simeq \frac{\tilde{m}}{\xi} \geq {\rm max} \left(T_{\rm GH}, T_{\rm max} \right) \ .
\label{eq:precond}
\ee
Throughout this section, we simplify our discussion by working in the instantaneous reheating limit, corresponding to a sufficiently large inflaton decay width. In such a limit there is no distinction between $T_{\rm max}$ and the reheating temperature $T_{\rm RH}$, and they are obtained from the energy scale of inflation
\be
T_{\rm RH} = T_{\rm max} = 0.66 \, g_*^{-1/4} \, E_I \ .
\ee
This provides an additional parameter, which we assume does not scan. With $E_I$ well below $M_{Pl}$, the boundary between Pre- and Post-Inflation scenarios is at $f \sim T_{RH}$.

In the absence of R-parity violation (RPV), the lightest supersymmetric particle (LSP) is stable and may contribute to the energy density of the universe. In our model the LSP is likely to be the axino, with mass given in \Eq{axinotree} and \Eq{axinoloop}. For large SUSY breaking scales this may be approximated as
\be
m_{\rm LSP} \sim \frac{\xi^2}{8\pi^2} \, \tilde{m} \, \sin 2\beta \ .
\label{eq:LSPmass}
\ee
For large values of the LSP mass
\be
m_{\rm LSP} \geq T_{\rm RH} \ ,
\label{eq:LSPmassCond}
\ee
inflation dilutes any primordial LSP abundance, and the plasma produced after reheating does not have enough thermal energy to produce LSP particles again. In such a region the dark matter is only made of axions, whereas if \Eq{eq:LSPmassCond} is not satisfied we have a two-component DM scenario. 

\begin{figure}
\begin{center}
\includegraphics[scale=.7]{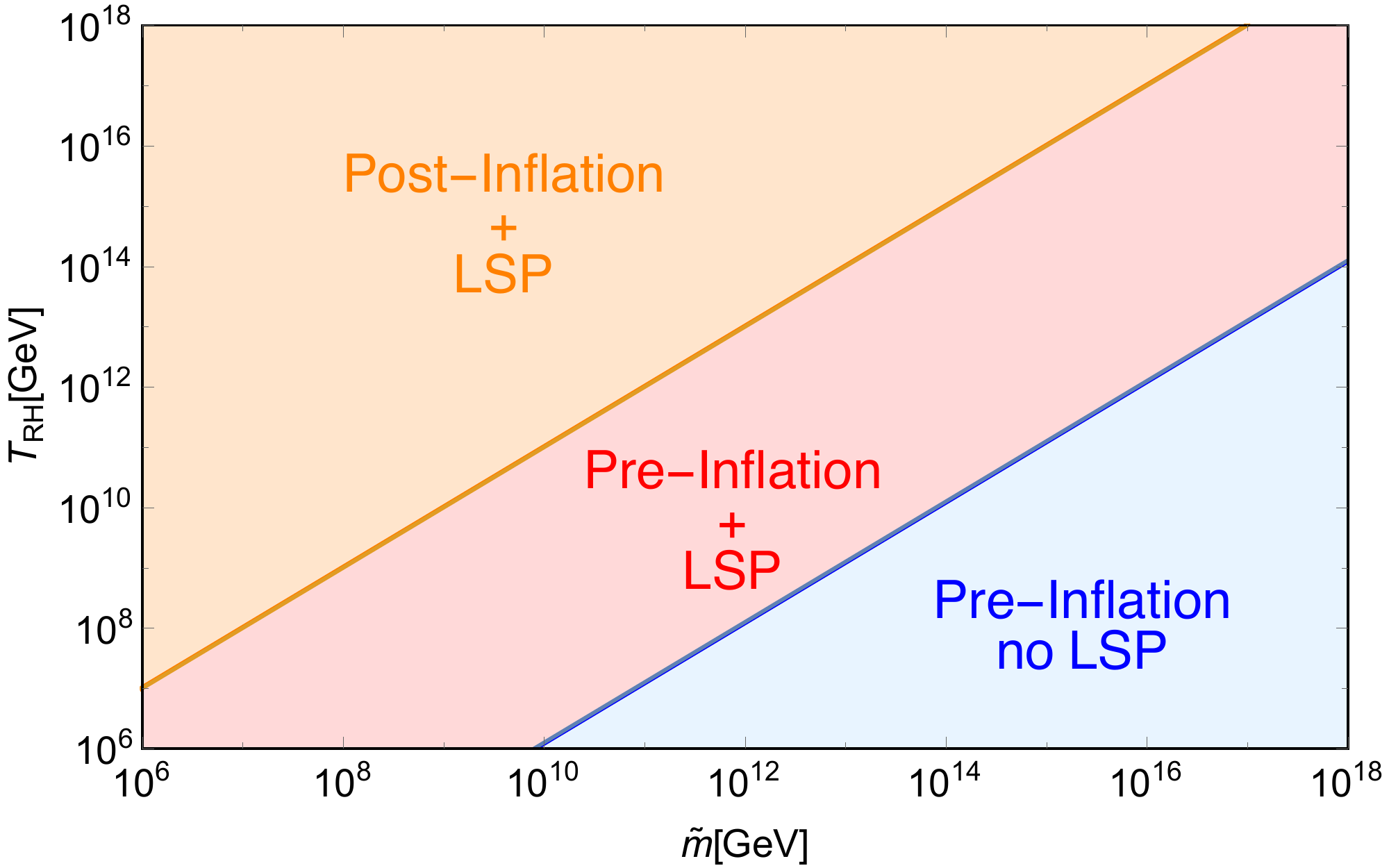}
\end{center}
\caption{Different cosmological histories in the $(\tilde{m}, T_{\rm RH})$ plane, assuming instantaneous reheating after inflation and taking $\xi = 0.1$.  Since $f \sim \tilde{m}/ \xi$, all regions contain axion dark matter.}
\label{fig:cosmology1}
\end{figure}

The parameter space in the $(\tilde{m}, T_{\rm RH})$ plane is sketched in \Fig{fig:cosmology1} for $\xi = 0.1$. In the orange region \Eq{eq:precond} is not satisfied, and therefore we are in the Post-Inflation scenario. Given our choice for $\xi$, points of the parameter space in the orange region never satisfy \Eq{eq:LSPmassCond}, thus we always have a LSP relic density. The red and blue regions are associated to the Pre-Inflation scenario, but only the red one has a contribution to the DM relic density from the LSP. Throughout the orange and red regions of \Fig{fig:cosmology1} the LSP abundance exceeds that of our universe by many orders of magnitude. Furthermore this region is anthropically excluded because the resulting dark matter density is so large that stellar systems suffer close encounters \cite{Tegmark:2005dy}. Hence the only region of interest in \Fig{fig:cosmology1} is the blue region, which has single component axion dark matter, and is the focus of the next sub-section.\footnote{For lower values of $\tilde{m}$ than shown in \Fig{fig:cosmology1}, the axino dark matter abundance is not catastrophic.  This part of parameter space does not describe our universe since $f \sim \tilde{m}/ \xi$ is too low; but it might allow observers, so we explore it in Section \ref{subsection:LSPDM}.}
This conclusion persists even if the axino mass is dominated by a contribution of order the gravitino mass, or if the gravitino is the LSP \cite{Cheung:2011mg}, although the size of the blue region may change.

\subsection{Pure Axion Dark Matter}
\label{subsection:axionDM}

We study in more detail the blue region of  \Fig{fig:cosmology1},
which has single component axion DM with relic density
\be
\rho_a = \rho_0 \; 1.67 \times \left[ \theta_i^2 + \left(\frac{H_I}{2 \pi f}\right)^2 \right] \, F_1(\theta_i) \, \left( \frac{f}{10^{12} \, {\rm GeV}} \right)^{1.2} \ .
\label{eq:axiondensity}
\ee
with $\rho_0$ the observed DM abundance. Since we are in the Pre-Inflation scenario, the initial misalignment angle $\theta_i$ is stretched by inflation to space-time regions beyond our current horizon, and therefore should not be averaged. Incidentally, in the blue region we do not need to worry about the domain wall problem (despite the fact that our model has domain wall number $N = 3$), since inflation dilutes the associated energy density. Finally, the function $F_1(\theta)$ has been evaluated numerically in Ref.~\cite{Bae:2008ue}. 

\begin{figure}
\begin{center}
\includegraphics[height=2.73in]{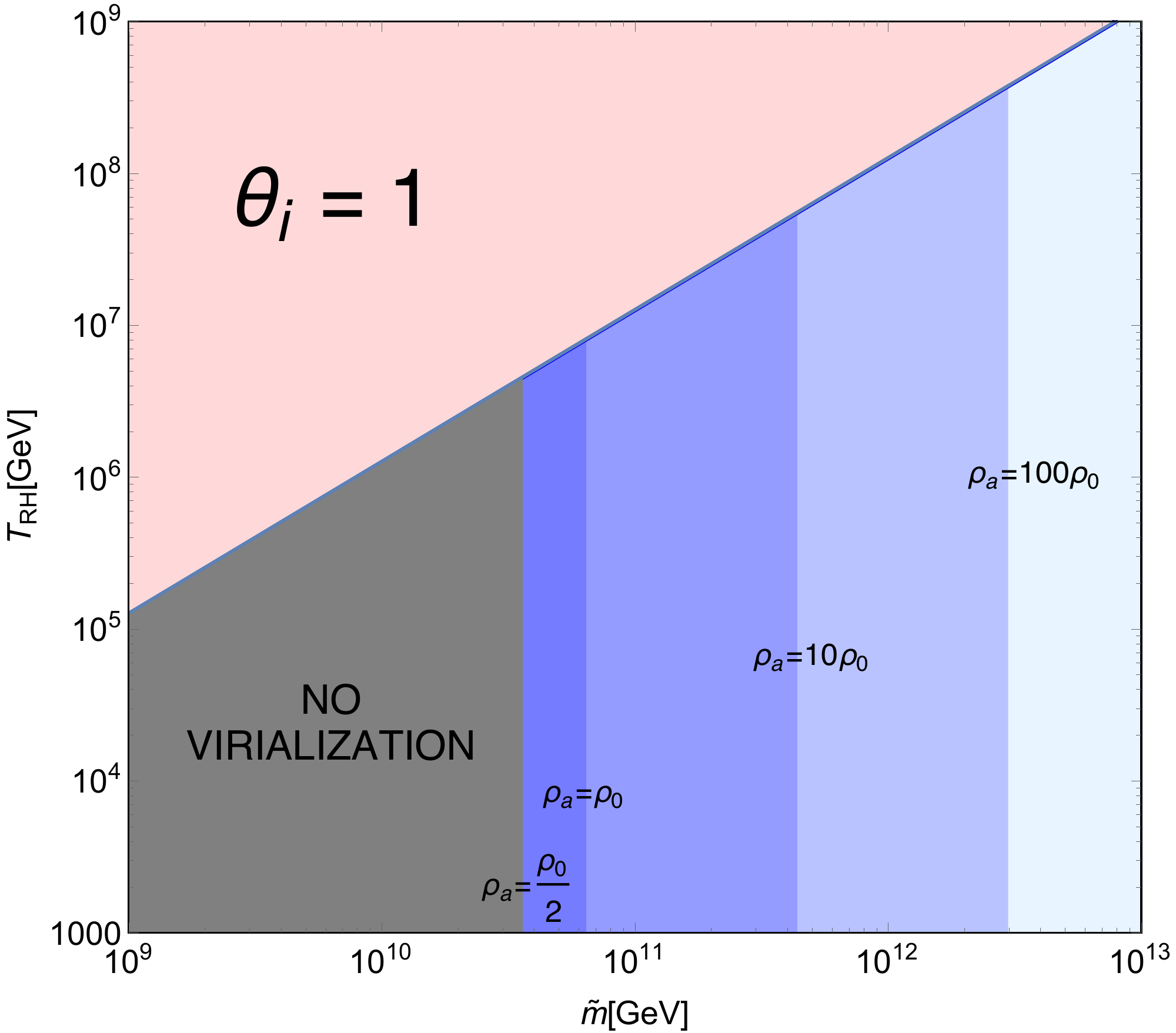} $\qquad$
\includegraphics[height=2.73in]{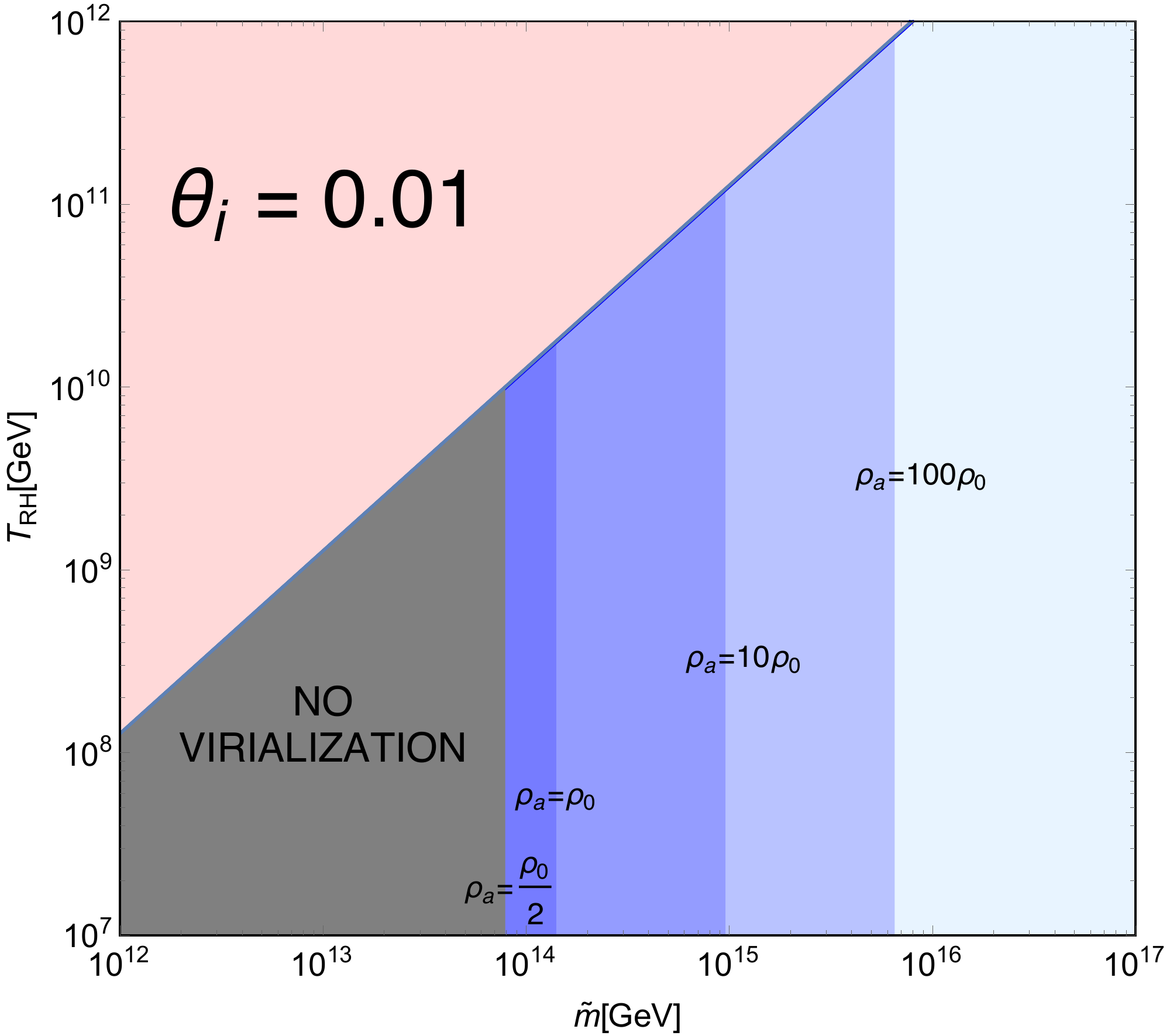} 
\end{center}
\caption{Contours of the axion dark matter density in the Pre-Inflation region with no LSP dark matter, with the initial misalignment angle fixed at 1 (0.01) in the left (right) panel. The gray region is excluded by the halo virialization requirement of \cite{D'Eramo:2014rna}, and the red region is excluded by extremely large LSP production during the instantaneous reheating.}
\label{fig:cosmology2}
\end{figure}

In  \Fig{fig:cosmology2} we show isocontours for the axion energy density for fixed values of the initial misalignment angle of 0.01 and 1 in the left and right panels.   The blue shaded regions have varying axion abundance, and we caution the reader that $\theta_i$ scans as well as $\tilde{m}$.  The gray region identifies the catastrophic virialization boundary discussed in Ref.~\cite{D'Eramo:2014rna}, while the pink region has catastrophic close stellar encounters if R parity is conserved.

As already mentioned at the beginning of this Section, we assume a power-law prior distribution for the SUSY breaking scale. Keeping in mind that we are only interested in the Pre-Inflation scenario we have the double differential probability distribution
\be
\frac{d P}{d \log \tilde{m} \; d \theta} \; \propto \; \theta\left(\rho_a - \rho_a^{\rm min}\right) \theta\left(\rho_a^{\rm max} - \rho_a\right)
\; \frac{v^2}{v^2 + \tilde{m}^2} \; \frac{1}{1 + \rho_a / \rho_B} \, \tilde{m}^n \ .
\label{eq:mtildeprob}
\ee
The first two theta functions describe catastrophic boundaries, associated to virialization and close encounters, respectively. In what follows, we assume $\rho_a^{\rm min} = 0.5 \rho_{D0}$ and $\rho_a^{\rm max} = 10^4 \rho_{D0}$. The factor involving $v$ accounts for the fine-tuning to satisfy the anthropic EWSB requirement. The term dependent on the ratio  $\rho_a/\rho_B$ is a measure factor~\cite{Freivogel:2008qc, Bousso:2013rda}, and $\rho_B$ is the baryon energy density which we assume does not scan. Finally, $n$ is an unknown parameter describing the distribution. 

The axion energy density $\rho_a$ depends on the three variables $(\theta_i, f, H_I)$, or equivalently on $(\theta_i , \tilde{m}, T_{\rm RH})$. Below a certain $\theta_{\rm min} \sim H_I / (2\pi f)$ the axion density is not dependent on $\theta_i$ anymore. The condition of no LSP relic density imposes
\be
\tilde{m} >  \frac{8 \pi^2}{\xi^2} T_{\rm RH} \ .
\label{eq:blueregion}
\ee
This gives a lower bound on the range of $\tilde{m}$ where we are allowed to scan for fixed $\xi$ and $T_{\rm RH}$. For example, for $\xi \simeq 0.1$ as chosen before, we end up with the condition $\tilde{m} > 8 \times 10^3 \, T_{\rm RH}$. For fixed reheating temperature, the $H_I$-dependent term in \Eq{eq:axiondensity} is a decreasing function of $\tilde{m}$, and when the inequality in \Eq{eq:blueregion} is saturated its size results in
\be
 \frac{H_I}{2 \pi f} \simeq \frac{1}{2 \pi} \; \frac{\xi}{\tilde{m}} \; 
\sqrt{\frac{8 \pi}{3}} \frac{1}{M_{\rm Pl}} \left( \frac{T_{\rm RH}}{0.66 \, g_*^{-1/4}} \right)^2  \leq 
0.013 \; \xi^3 \, g_*^{1/2} \, \frac{T_{\rm RH}}{M_{\rm Pl}} \ .
\ee
Given the condition in \Eq{eq:blueregion}, we do not want $T_{\rm RH}$ to be too high, otherwise we would have an enormous SUSY breaking scale. To be conservative we consider $T_{\rm RH}$ such that the condition in \Eq{eq:blueregion} becomes $\tilde{m} > 10^9 \, {\rm GeV}$ (or equivalently $f > 10^{10} \, \GeV$). This implies that the $H_I$-dependent contribution to the axion density in \Eq{eq:axiondensity} is completely negligible, as it becomes relevant for values of $\theta_{\rm min}$ that would require a transplanckian PQ breaking scale in order to satisfy the virialization boundary. 

\begin{figure}
\begin{center}
\includegraphics[height=2.4in]{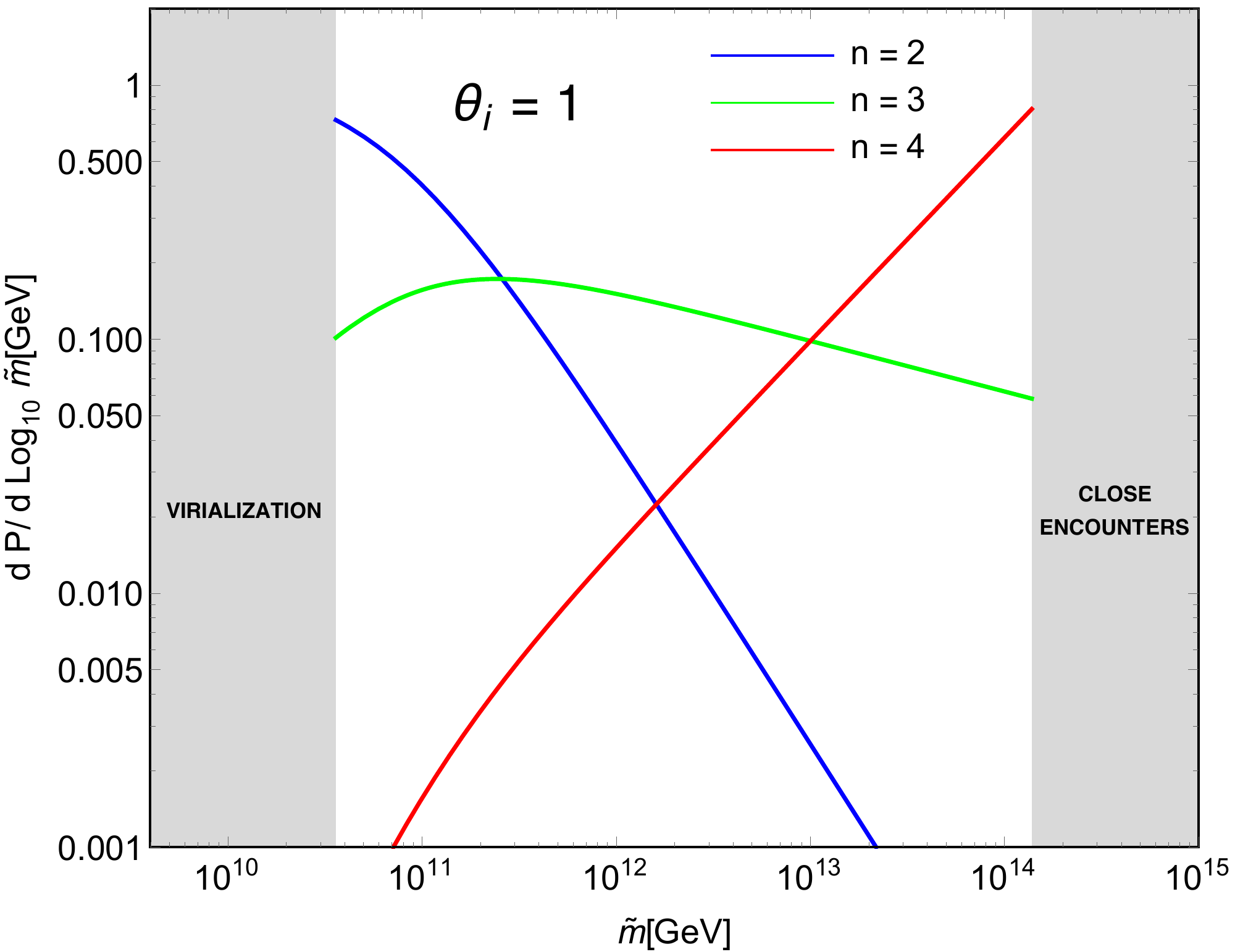} $\qquad$
\includegraphics[height=2.4in]{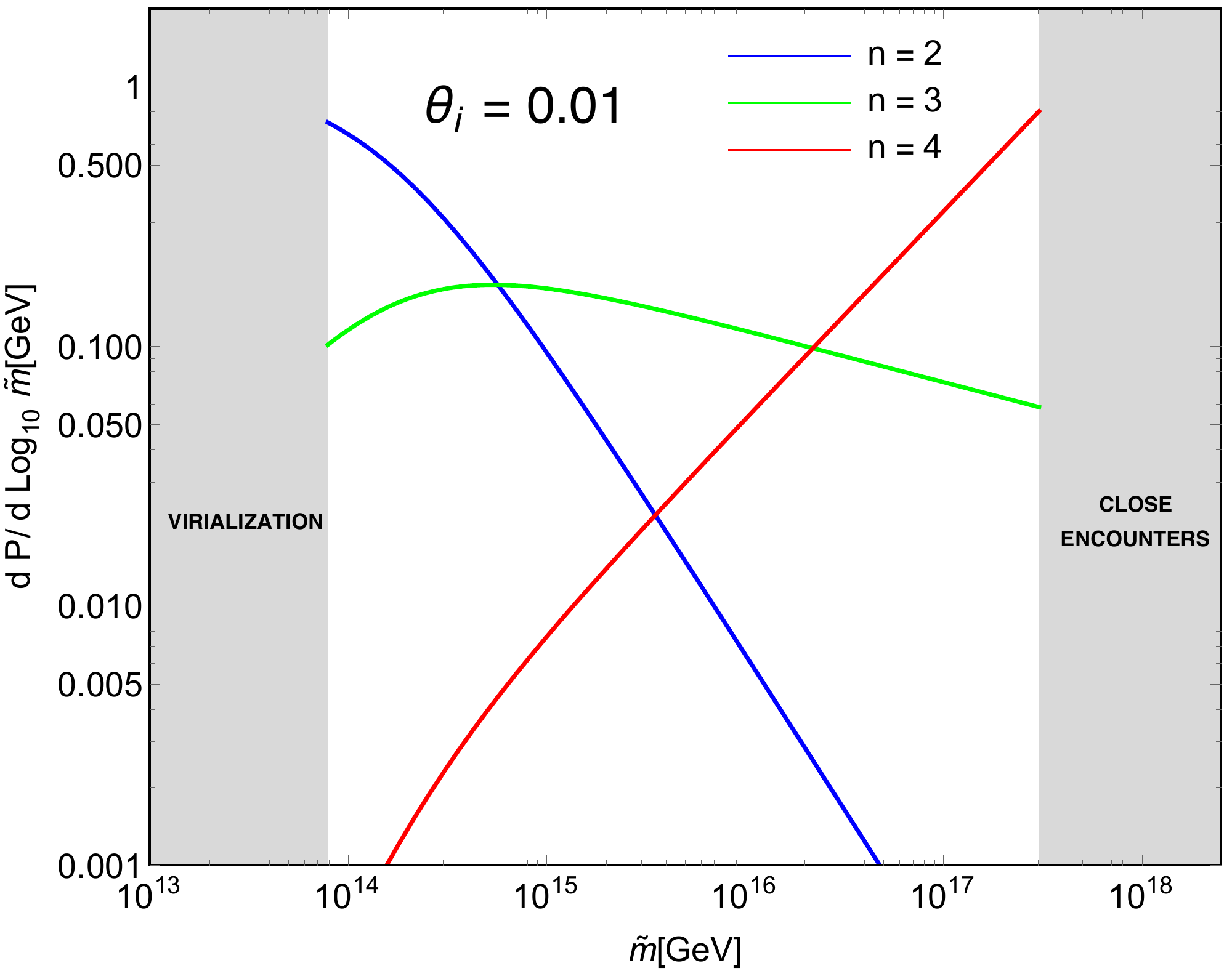} 
\end{center}
\caption{Probability distributions for $\tilde{m}$ for $\theta_i = 1$ (left panel) and $\theta_i = 0.01$ (right panel). In each case we show the probability distribution for three representative values of $n$.}
\label{fig:cosmology3}
\end{figure}

For simplicity we study two slices of the $(\tilde{m}, \theta_i)$ scanning parameters at fixed values of 
\be
\theta_i = \left\{ \begin{array}{cccccl}
10^{-2} & & & & & {\rm fine \; tuned \; angle} \\
1 & & & & & {\rm typical \; value} 
\end{array} \right. \ .
\ee
Fixing $\theta_i$ determines the allowed range for $f$ between the virialization and close encounters anthropic boundaries, and taking $f= \tilde{m}/\xi$ gives
\begin{align}
7.9 \times 10^{14} \, \GeV \leq \frac{\tilde{m}}{\xi} &\, \leq 3 \times 10^{18} \, \GeV   \, \qquad \qquad \qquad \theta_i = 10^{-2} \ , \\
3.6 \times 10^{11} \, \GeV \leq \frac{\tilde{m}}{\xi} & \, \leq 1.4 \times 10^{15} \, \GeV  \, \qquad \qquad \qquad \theta_i = 1 \ .
\end{align}

In \Fig{fig:cosmology3} we plot the normalized probability distributions for $\tilde{m}$ for $n=2,3,4$ for each of these slices of the multiverse and for $\xi = 0.1$.
The examples of $n=2,3,4$ shown by the red, green and blue curves illustrate three different multiverse scenarios for axion dark matter
\begin{itemize}
\item
For $n = 2$ the probability distribution for $\tilde{m}$ is peaked towards small values, leading to typical observers  close to the virialization boundary, as proposed in Ref.~\cite{D'Eramo:2014rna}. There is a caveat to this case, if $\tilde{m}$ is further decreased by many orders of magnitude to the weak scale, then LSP dark matter could satisfy the virialization requirements. This is addressed in the next sub-section. 
\item The case $n = 3$ features a peak in the probability distribution, as a consequence of the measure factor from Ref.~\cite{Freivogel:2008qc}.  This has the remarkable feature of explaining why the baryon and dark matter energy densities are comparable.
\item Finally, $n = 4$ has a probability distribution peaked at large values of $\tilde{m}$, so that typical observers are near the close encounter boundary.  The proximity of our universe to this boundary is possible, but has not been demonstrated.
\end{itemize}

For simplicity, above we studied fixed $\theta_i$ slices of the multiverse.  In fact one must study the probability distribution of \Eq{eq:mtildeprob} over the full range of scanning parameters $(\tilde{m}, \theta_i)$.  One again discovers the above three behaviors, but the corresponding values of $n$ are affected by the larger scan, in particular by the possibility of $\theta_i$ running to small values at large $f$.  For a distribution for $\tilde{m}$ that is not very steep, $n \leq 2$, one discovers that typical observers have $\theta_i$ order unity and values of $f$ of order $10^{11}$ GeV, close to the virialization boundary\cite{D'Eramo:2014rna}.  For small values of $n$, it is the cost of electroweak symmetry breaking that prefers values of $f$ and $\tilde{m}$ as low as allowed by virialization.
 
\subsection{The Irrelevance of LSP Dark Matter}
\label{subsection:LSPDM}

\begin{figure}
\begin{center}
\includegraphics[height=2.7in]{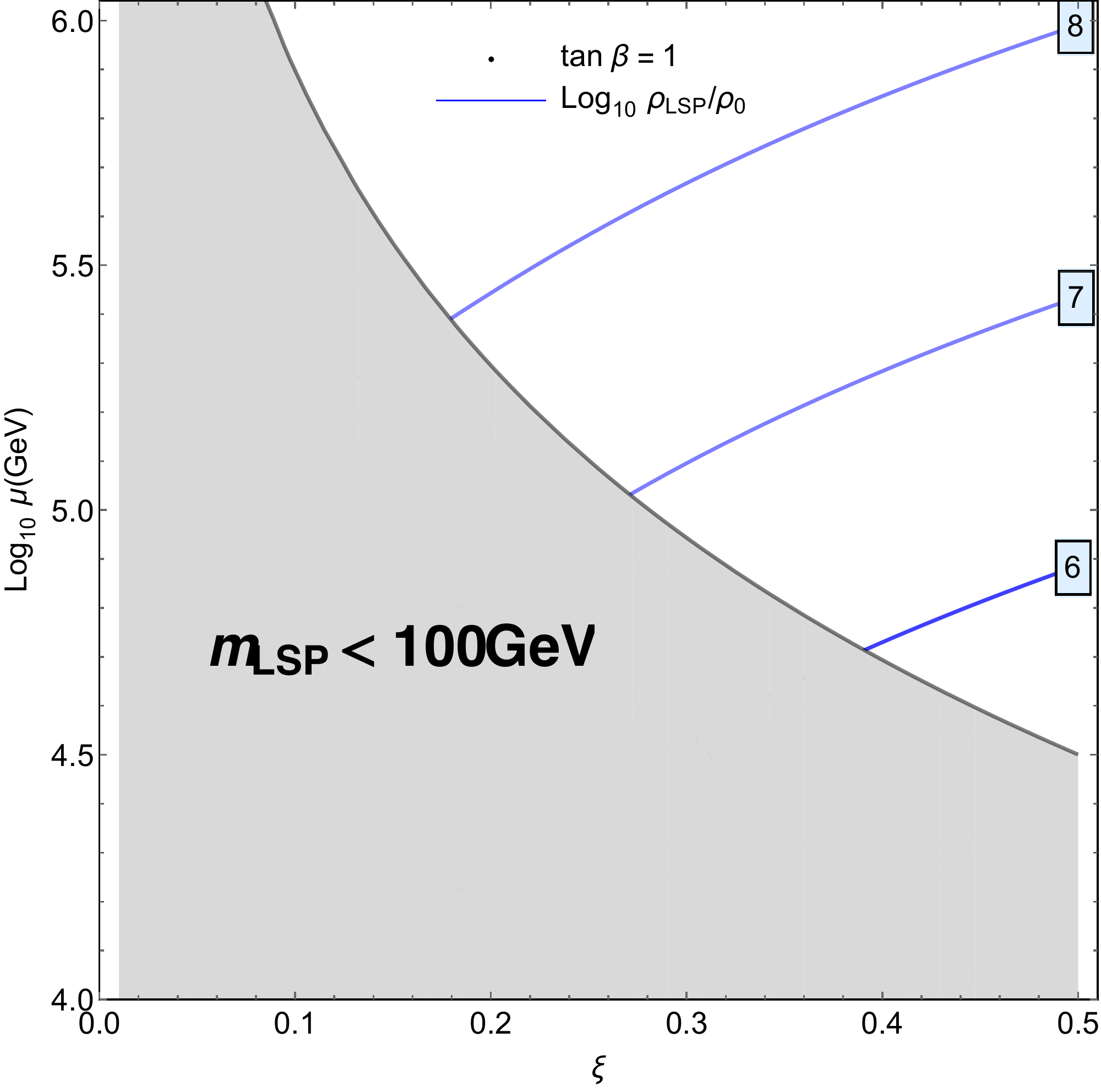} $\qquad$ \includegraphics[height=2.7in]{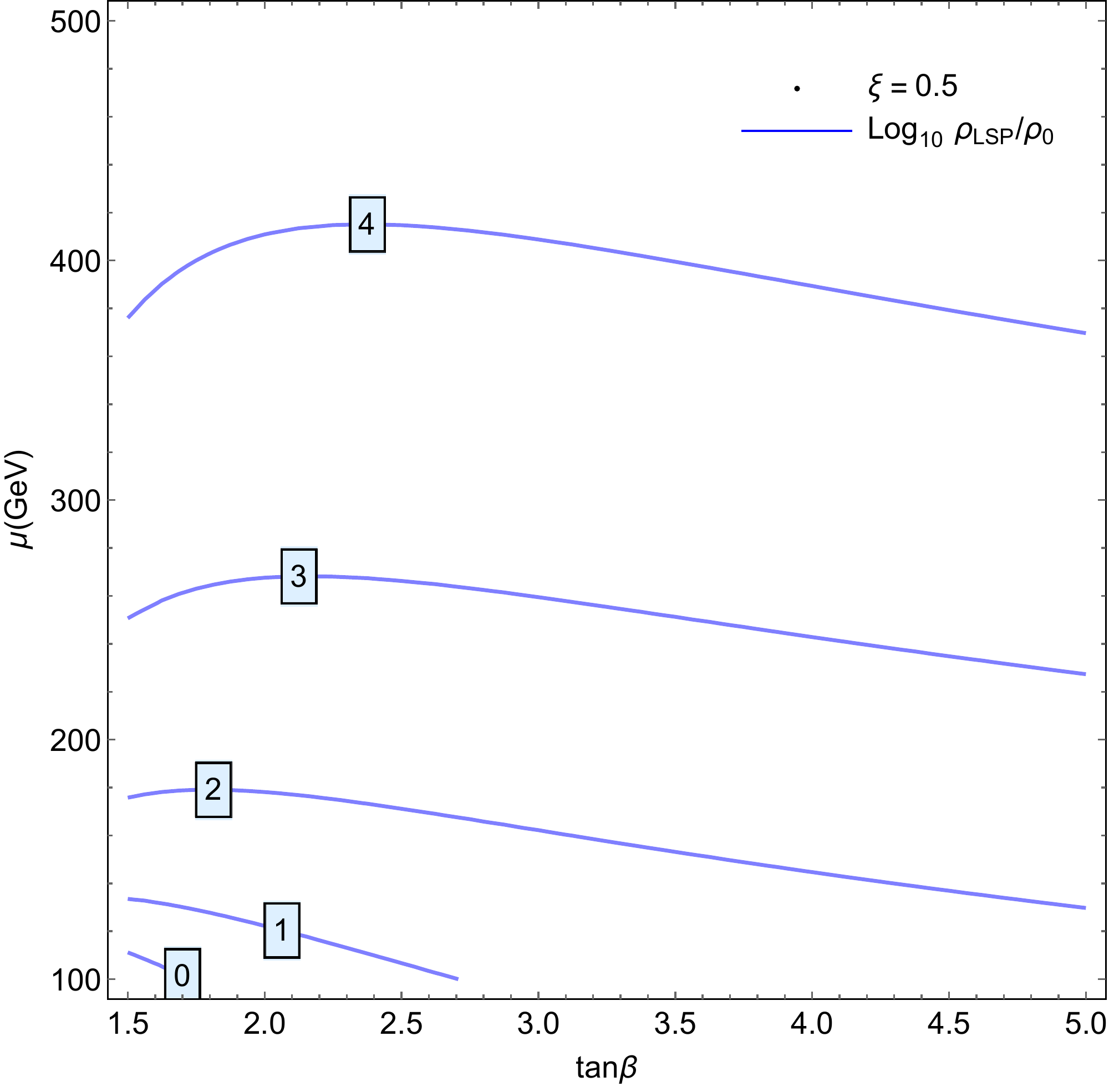}
\end{center}
\caption{LSP relic density for $\mu > 10 \; \TeV$ (left panel) and $\mu < 10 \; \TeV$ (right panel).}
\label{fig:freezeout}
\end{figure}

We conclude this Section with the results of the freeze-out calculation for the LSP relic density, justifying why we did not consider this contribution in the discussion above. The LSP is an admixture of the neutral fermions, which are the axino and the two neutral higgsinos. For large values of $\tilde{m} \sim \mu$, of the order of $10 \; \TeV$ or larger, the LSP is mostly along the axino direction and has a mass as given in \Eq{eq:LSPmass}. For such large values of the superpartner mass scale, we can neglect the mass splitting between the two higgsino-like states and integrate out the Dirac field with mass $\mu$. We find the effective Lagrangian
\be
\mathcal{L}_{ \tilde{a} \tilde{a}HH} =  - \frac{\xi^2}{2 \mu} \, \sin2\beta \, \tilde{a} \tilde{a}  \, h^\dagger h  + {\rm h.c.} \ ,
\label{eq:fermioneff}
\ee
where $h$ is the SM Higgs doublet and in this sub-section $\xi = \xi(\mu)$. This interaction mediates the annihilation process $\tilde{a} \tilde{a} \rightarrow h^\dag h$, which keeps $\tilde{a}$ in thermal equilibrium until freeze-out is achieved. The (thermally averaged) annihilation cross section reads
\be
\langle \sigma_{\tilde{a} \tilde{a} \rightarrow h^\dag h} v_{\rm rel} \rangle = \frac{\xi^4}{32 \pi \mu^2} \, \sin^22\beta \, \frac{T}{6 \, m_{\rm LSP}}  \ .
\ee
Since we consider large values of $\mu \sim \tilde{m}$ we are forced to pick $\tan\beta \sim 1$. A standard freeze-out calculation gives the LSP density isocontours shown in the left panel of \Fig{fig:freezeout}. The gray area identifies the parameter space region where $m_{\rm LSP} < 100 \; \GeV$ and therefore the annihilation to SM Higgses cannot control the freeze-out. As manifest from the figure, the LSP relic density is at least $6$ orders of magnitude above the one we observe, therefore this region is anthropically excluded by close encounters. 

This conclusion no longer holds for $\mu$ close to the weak scale. In this limit we recover the parameter space region of the singlet-doublet state studied in Refs.~\cite{Mahbubani:2005pt,D'Eramo:2007ga,Enberg:2007rp,Cohen:2011ec}, with the singlet mass not as a free parameter but rather one-loop suppressed. As is well known, this model has regions of well-tempered $\tilde{\chi}_{\rm LSP}$ where we reproduce the observed DM density. The one-loop suppression of the singlet mass is responsible for a very light LSP, always below $100 \, \GeV$. The freeze-out is controlled by annihilation to SM fermions, with (thermally averaged) cross section
\be
\langle \sigma_{\tilde{\chi}_{\rm LSP} \tilde{\chi}_{\rm LSP} \rightarrow \overline{f} f} v_{\rm rel} \rangle = 
\left( V_{13} V_{23} \right)^2 \frac{g^4 }{24 \pi c_w^4} \frac{m_{\rm LSP}^2}{\left( 4 m_{\rm LSP}^2 - m_Z^2\right)^2 + m_Z^2 \Gamma_Z^2}  \sum_f \left(g_{Vf}^{2}+g_{Af}^{2}\right) \, \frac{T}{6 \, m_{\rm LSP}}  \ .
\label{eq:sigmatofermions}
\ee
The sum over the vector $g_{Vf}$ and axial $g_{Af}$ couplings to the $Z$ boson runs over all the SM fermions except the top quark, and it reads
\be
\sum_f \left(g_{Vf}^{2}+g_{Af}^{2}\right) = 7.31 \ .
\ee
The suppression factor $(V_{13} V_{23})^2$ arises from the mass mixing (for details see Refs.~\cite{D'Eramo:2007ga,Enberg:2007rp}). We fix $\xi = 0.5$, since this parameter controls the mixing and we need to mix into doublet states in order to have efficient annihilations. In this case $\tan\beta$ is not bounded to be close to $1$ anymore, since all the parameters are at the weak scale. Moreover, for $\tan\beta = 1$ the mixing factor $(V_{13} V_{23})^2 = 0$ as a consequence of a custodial symmetry of the model in this limit, therefore we have to be away from this limit. We show our results in the right panel of \Fig{fig:freezeout}, where we plot isocontours for the LSP relic density in the $(\tan\beta, \mu)$ plane. We observe that there are anthropically viable regions, and even a line where the observed abundance is reproduced. 

Hence, regions of the multiverse with $\tilde{m}$ of order the weak scale have amounts of dark matter from axino freeze-out that would allow observers.  Furthermore, a multiverse distribution  with $n<2$ would make such regions with natural electroweak symmetry breaking more probably than regions with much larger $\tilde{m}$.  However, there are several possibilities for avoiding this unrealistic situation.  For example, if the reheat temperature after inflation were above about 100 TeV the production of LSP axinos or gravitinos by thermal scattering for freeze-in can lead to excessive amounts of dark matter \cite{Cheung:2011mg}.  Alternatively, for  $n<2$, the simple power law of $\tilde{m}^n$ contained in \Eq{eq:mtildeprob} may apply only for a limited range of $\tilde{m}$ near the intermediate scale, breaking down well before TeV-scale values are reached.

\section{Generality of the Higgs Mass Prediction}
\label{section:generality}
 We have presented our prediction for the Higgs boson mass in a simple supersymmetric axion model.  In fact the prediction has a much wider generality.
 
Suppose the SM is valid to some high scale $\Lambda > 10^{10}$ GeV, and that the theory at this scale involves a real SM singlet scalar $s$ with a scalar potential
 \be
 V(h,s)= \lambda_h \, (h^\dagger h)^2 +
\frac{m_s^2}{2} s^2+A \, s h^\dagger h+ \dots
 \ee
with $m_s \sim \Lambda$.  On integrating out $s$, the SM Higgs quartic in the low energy effective theory at scale $m_s$ becomes
\be
\lambda = \lambda_h - \frac{A^2}{2 m_s^2}.
\label{eq:lambdaSM}
\ee
An electroweak vacuum with lifetime greater than $10^{10}$ years requires $\lambda > \lambda_{\textrm{cr}}$, so we must insist that $\lambda_h > \lambda_{\textrm{cr}}$. 

Our Higgs mass prediction results from two assumptions. 
First, at the scale $\Lambda$, $\lambda_h$ is typically smaller than $A^2/m_s^2$, so that typically $\lambda <  \lambda_{\textrm{cr}}$.
Second, $A$ has several contributions with at least one scanning in the multiverse so that there is the possibility of cancellations among these contributions, giving a probability distribution for $A$ of $dP \propto dA$ at small $A$ and the distribution shown in Figure \ref{fig:probintro}.  With these assumptions, the Higgs mass prediction is as in Figure \ref{fig:ProbabilityLambda3}, with $\mu_c^2$ a 1-loop factor larger than $\Lambda^2$ and $\lambda_h$ parameterized by $\xi$ and $\tan \beta$ as in eq. (\ref{quartich}).


\section*{Acknowledgments}
We thank Raymond Co and Satoshi Shirai for useful discussions. This work was supported in part by the Director, Office of Science, Office of High Energy and Nuclear Physics, of the US Department of Energy under Contract DE-AC02-05CH11231 and by the National Science Foundation under grants PHY-1002399 and PHY-1316783. F.D. is supported by the Miller Institute for Basic Research in Science.

\appendix

\section{RG Equations}

In this Appendix we collect the one-loop RG equations for our model. We start from the one-loop running of the gauge couplings, whose evolution is governed by 
\be
\mu \frac{d g}{d \mu} \equiv \beta(g) \ , \qquad \qquad \qquad \beta(g) \equiv \frac{g^3}{(4 \pi)^2} \,  b^{(1)} \ .
\ee
We use the $SU(5)$-normalized $g_1$, related to the SM hypercharge by $g_1 = \sqrt{5/3} g^\prime$. We consider a degenerate spectrum, with all superpartners at $\mu_{\rm SUSY} = \tilde{m}$. The solutions are
\be
\frac{1}{\alpha_a(\mu)} = \frac{1}{\alpha_a(m_Z)} - \frac{1}{2\pi} 
\left[ b_a^{(1) {\rm SM}} \ln\left(\frac{\mu}{m_Z}\right) 
+ \theta\left(\mu - \tilde{m} \right) \Delta b_a^{(1) {\rm MSSM}}  \ln\left(\frac{\mu}{\tilde{m}}\right) \right] \ ,
\label{eq:gaugecouplingrunning}
\ee
with coefficients 
\be
b^{(1) {\rm SM}} =(41/10,\,-19/6,\,-7),\qquad \Delta b^{(1) {\rm MSSM}} =(5/2,\,25/6,\,4).
\label{eq:betacoeff}
\ee

The RG evolution of the superpotential couplings has only contributions from the  wave-function renormalization, as a consequence of the supersymmetric non-renormalization theorem~~\cite{Salam:1974jj,Grisaru:1979wc,Seiberg:1993vc}. The RG equations for the top Yukawa $y_t$ and $\xi$ read
\begin{align}
\mu \frac{d \xi}{d \mu} =  & \, \frac{\xi}{16\pi^2} \left[ 4 \xi^2 + 3 y_t^2 - 3 g_2^2 - \frac{3}{5} g_1^2 \right] \ , \\ 
\mu \frac{d y_t}{d \mu} = & \, \frac{y_t}{16\pi^2} \left[ \xi^2 + 6 y_t^2 - \frac{16}{3} g_3^2 - 3 g_2^2  -  \frac{13}{15} g_1^2 \right] \ .
\label{eq:runningW}
\end{align}
The top Yukawa is present also when the heavy degrees of freedom are integrated out, and we know its boundary condition at the weak scale. For the SM matter field content, the top Yukawa RG evolution reads~\cite{Machacek:1983fi}
\be
\left. \mu \frac{d y_t}{d \mu}\right|_{\rm SM} = \frac{y_t}{16\pi^2} \left[ \frac{9}{2} y_t^2 - 8 g_3^2 - \frac{9}{4} g_2^2 - \frac{17}{20} g_1^2 \right] \ .
\ee
The top Yukawa top is matched at the SUSY breaking scale as follows
\be
\left. y_t(\tilde{m}) \right|_{\rm MSSM} = \frac{\sqrt{1 + \tan^2\beta}}{\tan\beta} \left. y_t(\tilde{m}) \right|_{\rm SM} \ .
\label{eq:ytopmatching}
\ee

Among the RG equations for the soft terms, the ones for the gaugino masses are the only ones which can be analytically solved. The RG equations read~\cite{Ellwanger:2009dp}
\be
\frac{d M_a}{d \log \mu} = \frac{g_a^3\, b_a}{16\pi^2} \, \frac{2 M_a}{g_a} \ ,
\ee
which implies $M_a g_a^{-2} = {\rm const}$.  The RG equations for the A-terms result in
\begin{align}
\mu \frac{d A_\xi}{d \mu} = & \, \frac{1}{16 \pi^2} \left[ 8 A_\xi \xi^2 + 6 A_t y^2_t + 6 g_2^2 M_2 + \frac{6}{5} g_1^2 M_1 \right]\ , \\ 
\mu \frac{d A_t}{d \mu} = & \,  \frac{1}{16 \pi^2} \left[ 12 A_t y_t^2 + 2 A_\xi \xi^2 + \frac{32}{3} g_3^2 M_3 + 6 g_2^2 M_2  + \frac{26}{15} g_1^2 M_1 \right] \ .
\end{align}
Finally, the RG equations for the scalar soft masses read~\cite{Ellwanger:2009dp}
\begin{align}
\frac{d m^2_{H_u}}{d \log\mu} = & \,  \frac{1}{16\pi^2} \left[ 2 \xi^2 \left(m^2_{H_u} + m^2_{H_d} + m^2_S + A_\xi^2 \right)  + 6 y_t^2  \left(m^2_{H_u} + m^2_{Q} + m^2_u + A_t^2 \right) + \right. \\ & \left. \quad\qquad - 6 g_2^2 M_2^2 - \frac{6}{5} g_1^2 M_1^2 + \frac{3}{5} g_1^2 \mathcal{S}  \right] \ , \\
\frac{d m^2_{H_d}}{d \log\mu} = & \,  \frac{1}{16\pi^2} \left[ 2 \xi^2 \left(m^2_{H_u} + m^2_{H_d} + m^2_S + A_\xi^2 \right) - 6 g_2^2 M_2^2 - \frac{6}{5} g_1^2 M_1^2 - \frac{3}{5} g_1^2 \mathcal{S}  \right] \ , \\
\frac{d m^2_{S}}{d \log\mu} = & \,  \frac{1}{16\pi^2} \left[ 4 \xi^2 \left(m^2_{H_u} + m^2_{H_d} + m^2_S + A_\xi^2 \right) \right] \ , \\
\frac{d m^2_{Q}}{d \log\mu} = & \,  \frac{1}{16\pi^2} \left[ 2 y_t^2 \left(m^2_{H_u} + m^2_{Q} + m^2_u + A_t^2  \right) - \frac{32}{3} g_3^2 M_3^2 - 6 g_2^2 M_2^2 - \frac{2}{15} g_1^2 M_1^2 + \frac{1}{5} g_1^2 \mathcal{S} \right] \ , \\
\frac{d m^2_{u}}{d \log\mu}  = & \,  \frac{1}{16\pi^2} \left[ 4 y_t^2 \left(m^2_{H_u} + m^2_{Q} + m^2_u + A_t^2 \right) - \frac{32}{3} g_3^2 M_3^2 - \frac{32}{15} g_1^2 M_1^2 - \frac{4}{5} g_1^2 \mathcal{S} \right]  \ ,
\end{align}
where we have introduced
\be
\mathcal{S} = {\rm Tr} \left[Y_i m^2_{i}\right]  \simeq m^2_{H_u} - m^2_{H_d} + m_Q^2 - 2 m_u^2 - 2 m_d^2 \ .
\ee

\section{RG Analytical Solution in a Simplified Case}

We consider a simplified case of our model, where we neglect the top Yukawa, the top A-term, the gauge couplings and the soft masses. The one-loop RG equations system reads
\begin{align}
\frac{d \xi}{d \log\mu} = & \,\frac{ 4 \xi^3}{16\pi^2}  \ , \\
\frac{d A_\xi}{d \log\mu}  = & \, \frac{8 A_\xi \xi^2}{16 \pi^2} \ , \\
\frac{d m^2_{H_u}}{d \log\mu} = & \, \frac{2 \xi^2}{16\pi^2} \left(m^2_{H_u} + m^2_{H_d} + m^2_S + A_\xi^2 \right)\ , \\
\frac{d m^2_{H_d}}{d \log\mu} = & \,  \frac{2 \xi^2}{16\pi^2} \left(m^2_{H_u} + m^2_{H_d} + m^2_S + A_\xi^2 \right) \ , \\
\frac{d m^2_{S}}{d \log\mu} = & \, \frac{4 \xi^2}{16\pi^2} \left(m^2_{H_u} + m^2_{H_d} + m^2_S + A_\xi^2 \right) \ .
\end{align}
This system can be completely solved analytically. We give initial conditions at the cutoff scale $M_*$, and we denote couplings at the cutoff with a star subscript (e.g. $\xi(M_*) = \xi_*$).

The running Yukawa coupling $\xi$ and A-term $A_\xi$ result in
\begin{align}
\xi(\mu) = & \, \frac{\xi_*}{\left[1 + \frac{\xi_*^2}{2 \pi^2} \log\left(\frac{M_*}{\mu}\right) \right]^{1/2}} \ , \\
A(\mu) =  & \, \frac{A_*}{1 + \frac{\xi^2_*}{2\pi^2} \log\left(\frac{M_*}{\mu}\right) } \ .
\end{align}
The remaining equations for the soft masses can be rewritten in a matrix form
\be
\frac{d}{d \log \mu} \left(\begin{array}{c} m^2_{H_u} \\ m^2_{H_d} \\ m^2_{S} \end{array} \right) = 
\frac{\xi(\mu)^2}{8 \pi^2} \left(\begin{array}{ccc} 1 & 1 & 1  \\ 1 & 1 & 1  \\ 2 & 2 & 2 \end{array} \right)
\left(\begin{array}{c} m^2_{H_u} \\ m^2_{H_d} \\ m^2_{S} \end{array} \right) + 
\frac{\xi(\mu)^2 A_\xi(\mu)^2 }{8 \pi^2}  \left(\begin{array}{c} 1 \\ 1 \\ 2 \end{array} \right) \ .
\ee
It is convenient to rotate the soft masses to another basis where the equations are decoupled. This is achieved by the following transformation
\be
\left(\begin{array}{c} Y_1 \\ Y_2 \\ Y_3 \end{array} \right)  = 
 \left(\begin{array}{ccc} \frac{\sqrt{3}}{2\sqrt{2}} & \frac{\sqrt{3}}{2\sqrt{2}} & \frac{\sqrt{3}}{2\sqrt{2}}  \vspace{0.1cm} \\ 
- \frac{1}{\sqrt{2}} & - \frac{1}{\sqrt{2}} & \frac{1}{\sqrt{2}}  \vspace{0.1cm} \\ 
- \frac{1}{2 \sqrt{2}} & \frac{3}{2 \sqrt{2}} & - \frac{1}{2 \sqrt{2}} \end{array} \right)  \left(\begin{array}{c} m^2_{H_u} \\ m^2_{H_d} \\ m^2_{S} \end{array} \right).
\ee
The equations for $Y_i$ variables are decoupled
\begin{align}
\frac{d Y_1}{d \log\mu} = & \, \frac{\xi(\mu)^2 }{2 \pi^2} Y_1 + \sqrt{6} \frac{\xi(\mu)^2 A_\xi(\mu)^2 }{8 \pi^2} \ ,  \\
\frac{d Y_2}{d \log\mu} = & \, 0 \ ,\\  
\frac{d Y_3}{d \log\mu} = & \, 0 \ ,
\end{align}
with solutions
\begin{align}
Y_1(\mu) = & \, \frac{C_1}{1 + \frac{\xi^2_*}{2\pi^2} \log\left(\frac{M_*}{\mu}\right) } +  \frac{\sqrt{6}}{4} \frac{A^2_*}{\left(1 + \frac{\xi^2_*}{2\pi^2} \log\left(\frac{M_*}{\mu}\right)\right)^2 } \ , \\  
Y_2(\mu) =  & \, C_2 \ , \\  
Y_3(\mu) = & \,  C_3 \ . 
\end{align}

The integration constants $C_i$ are set by the UV boundary conditions for the soft masses
\begin{align}
C_1 = & \,  \frac{\sqrt{6}}{4} \left( m^2_{H_u*} + m^2_{H_d*} + m^2_{S*}  - A_*^2 \right)   \ , \\
C_2 = & \, \frac{1}{\sqrt{2}} \left( - m^2_{H_u*} - m^2_{H_d*} + m^2_{S*}  \right) \ , \\
C_3 = & \, \frac{1}{2\sqrt{2}} \left( - m^2_{H_u*} + 3 m^2_{H_d*} - m^2_{S*}  \right) \ .
\label{eq:Ci}
\end{align}
We rotate back to the original basis, and we find the solution for the running soft masses
\begin{align}
m^2_{H_u}(\mu) = & \, \frac{1}{\sqrt{6}} \frac{C_1}{1 + \frac{\xi^2_*}{2\pi^2} \log\left(\frac{M_*}{\mu}\right) } +  \frac{1}{4} \frac{A^2_*}{\left(1 + \frac{\xi^2_*}{2\pi^2} \log\left(\frac{M_*}{\mu}\right)\right)^2 } - \frac{1}{\sqrt{2}} ( C_2 + C_3) \ , \\
m^2_{H_d}(\mu) = & \, \frac{1}{\sqrt{6}} \frac{C_1}{1 + \frac{\xi^2_*}{2\pi^2} \log\left(\frac{M_*}{\mu}\right) } +  \frac{1}{4} \frac{A^2_*}{\left(1 + \frac{\xi^2_*}{2\pi^2} \log\left(\frac{M_*}{\mu}\right)\right)^2 } + \frac{1}{\sqrt{2}}  C_3 \ , \\
m^2_{S}(\mu) = & \, \frac{2}{\sqrt{6}} \frac{C_1}{1 + \frac{\xi^2_*}{2\pi^2} \log\left(\frac{M_*}{\mu}\right) } +  \frac{1}{2} \frac{A^2_*}{\left(1 + \frac{\xi^2_*}{2\pi^2} \log\left(\frac{M_*}{\mu}\right)\right)^2 } + \frac{1}{\sqrt{2}}  C_2 \ . \\
\end{align}

\section{Anomalies of the PQ symmetry}\label{anomaliesPQ}
The lagrangian contains $\theta$-terms for the various gauge group. We normalize them as
\be
-\theta_3\frac{g_3^2}{32\pi^3}G^a_{\mu\nu}\tilde G_{\mu\nu}^a-\theta_2\frac{g_2^2}{32\pi^3}W^a_{\mu\nu}\tilde W_{\mu\nu}^a-\theta_1\frac{g_1^2}{32\pi^3}B_{\mu\nu}\tilde B_{\mu\nu}.
\ee
Under a chiral rotation
\be
\psi\to e^{i\alpha}\psi
\ee
where $\psi$ is one of the two component Weyl fermions of the model
\be
Q,~u^c,~d^c,~L,~e^c,~\tilde h_1,~\tilde h_2,~\tilde s,
\ee
the various $\theta$-terms transform as
\be
\theta\to \theta-2 T(r_\psi)\alpha
\ee
where $T(r_\psi)$ is the Dynkin index of the representation $r_\psi$. $T(r_\psi)=1/2$ for fundamental representations of $SU(N)$ and $T(r_\psi)=Y_\psi^2$ if the fermion $\psi$ has charge $Y_\psi$ under an abelian gauge group. For our model we get
\bea\label{PQshift1}
\theta_3&\to& \theta_3-N_g(2 Q_Q+Q_{u^c}+Q_{d^c})\alpha\\
\theta_2&\to&\theta_2-N_g(3Q_Q+Q_L)\alpha-(Q_1+Q_2)\alpha\\
\theta_1&\to&\theta_1-2N_g\left(6\left(\tfrac{1}{6}\right)^2Q_Q+3\left(\tfrac{2}{3}\right)^2Q_{u^c}+3\left(\tfrac{1}{3}\right)^2Q_{d^c}+2\left(\tfrac{1}{2}\right)^2Q_L+Q_{e^c}\right)\alpha+\\
&&-2\left(2\left(\tfrac{1}{2}\right)^2Q_1+2\left(\tfrac{1}{2}\right)^2Q_2\right)\alpha
\eea
where $N_g=3$ is the number of generations and the various color and weak-isospin multiplicities are easily understood. Under the assumption that the PQ symmetry is exact up to anomalies we obtain the following relations
\be\label{PQrel}
0=Q_Q+Q_{u^c}+Q_2=Q_Q+Q_{d^c}+Q_1=Q_L+Q_{e^c}+Q_1=Q_1+Q_2+Q_S.
\ee
Using these eq.~(\ref{PQshift1}) can be simplified
\bea
\theta_3&\to&\theta_3-N_gQ_S\alpha\\
\theta_2&\to&\theta_2-N_g(3Q_Q+Q_L)\alpha+Q_S\alpha\\
\theta_1&\to&\theta_1+N_g(3Q_Q+Q_L)\alpha+\left(1-\tfrac{8}{3}N_g\right)Q_S\alpha.
\eea
Defining
\be
-\theta_{EM}\frac{e^2}{32\pi^2} F_{\mu\nu}\tilde F_{\mu\nu},~~~ A_\mu=\sin\theta_W W_\mu^3+\cos\theta_W B_\mu
\ee
\be
\theta_{EM}\equiv \theta_1+\theta_2\to \theta_{EM}+\left(2-\tfrac{8}{3}N_g\right)Q_S\alpha
\ee
which depends only on the combination $Q_S\alpha$.

\section{Axion interactions}\label{axioninteractions}

To discuss the axion interactions at low energy it is convenient to perform a field redefinition to go to a basis in which all the matter fields are invariant under a PQ transformation. If $Q_\psi$ is the charge of the field $\psi$ this is obtained by
\be\label{PQrot}
\psi\to e^{i Q_\psi \frac{a}{f}}\psi.
\ee
Through its anomaly this rotation induces an interaction of the axion with the QCD and electromagnetic dual field strengths (see Appendix~\ref{anomaliesPQ})
\be\label{axiontoFFdual}
N_g\frac{g_3^2}{32\pi^2} \frac{a}{f}G_{\mu\nu}^a\tilde G_{\mu\nu}^a+\left(\frac{8}{3}N_g-2\right)\frac{e^2}{32\pi^2} \frac{a}{f}F_{\mu\nu}\tilde F_{\mu\nu}.
\ee
$N_g=3$ is the number of SM generations. Since $a$ transforms in the canonical way eq.~(\ref{shift}), the factor $N_g$ identifies the domain-wall number of the model which is thus different than unity. Eq.~(\ref{axiontoFFdual}) holds all the way down to the QCD confinement scale. This holds because in the basis we are using all the matter fields are neutral under the PQ rotation and they do not induce any modification in eq.~(\ref{axiontoFFdual}) as they are integrated out along the RG flow.

When the field redefinition in eq.~(\ref{PQrot}) is applied to the kinetic terms of the light fermions the following derivative interactions are obtained
\be\label{axiontofermions}
Q_S \left( \frac{\sin^2 \beta}{2}\frac{\partial_\mu a}{f} \, \bar e \gamma_\mu\gamma_5 e \; + \; \frac{\sin^2 \beta}{2}\frac{\partial_\mu a}{f} \, (\bar d \gamma_\mu\gamma_5 d+\bar s \gamma_\mu\gamma_5 s) \; + \; \frac{\cos^2 \beta}{2}\frac{\partial_\mu a}{f} \, \bar u \gamma_\mu\gamma_5 u \right).
\ee
We neglected derivative couplings of the axion to the fermionic vector currents. We notice that no interaction of the axion with the light Higgs current $iH^\dagger D_\mu H-iD_\mu H^\dagger  H$ is generated. Similarly to eq.~(\ref{axiontoFFdual}) the couplings in eq.~(\ref{axiontofermions}) are not renormalized as the various heavy fields are integrated out.


\begin{thebibliography}{0}
  
  
  \bibitem{Weinberg:1987dv}
S.~Weinberg,
Phys.\ Rev.\ Lett.\  {\bf 59}, 2607 (1987).

\bibitem{Martel:1997vi} 
  H.~Martel, P.~R.~Shapiro and S.~Weinberg,
  Astrophys.\ J.\  {\bf 492}, 29 (1998)
  [astro-ph/9701099].
  
\bibitem{Bousso:2007kq} 
  R.~Bousso, R.~Harnik, G.~D.~Kribs and G.~Perez,
  Phys.\ Rev.\ D {\bf 76}, 043513 (2007)
  [hep-th/0702115 [HEP-TH]].

\bibitem{Bousso:2010zi} 
  R.~Bousso, B.~Freivogel, S.~Leichenauer and V.~Rosenhaus,
  Phys.\ Rev.\ Lett.\  {\bf 106}, 101301 (2011)
  [arXiv:1011.0714 [hep-th]].

\bibitem{Bousso:2010im} 
  R.~Bousso, B.~Freivogel, S.~Leichenauer and V.~Rosenhaus,
  Phys.\ Rev.\ D {\bf 84}, 083517 (2011)
  [arXiv:1012.2869 [hep-th]].

  \bibitem{Agrawal:1997gf} 
V.~Agrawal, S.~M.~Barr, J.~F.~Donoghue and D.~Seckel,
Phys.\ Rev.\  D {\bf 57}, 5480 (1998)
[arXiv:hep-ph/9707380].

\bibitem{Hall:2014dfa} 
  L.~J.~Hall, D.~Pinner and J.~T.~Ruderman,
  arXiv:1409.0551 [hep-ph].

\bibitem{Peccei:1977hh} 
  R.~D.~Peccei and H.~R.~Quinn,
  Phys.\ Rev.\ Lett.\  {\bf 38}, 1440 (1977).
  
\bibitem{Weinberg:1977ma} 
  S.~Weinberg,
  Phys.\ Rev.\ Lett.\  {\bf 40}, 223 (1978).
  
\bibitem{Wilczek:1977pj} 
  F.~Wilczek,
  Phys.\ Rev.\ Lett.\  {\bf 40}, 279 (1978).
  
\bibitem{Feldstein:2004xi} 
  B.~Feldstein, L.~J.~Hall and T.~Watari,
  Phys.\ Lett.\ B {\bf 607}, 155 (2005)
  [hep-ph/0411013].
  
\bibitem{Frieman:1987ui} 
  J.~A.~Frieman, S.~Dimopoulos and M.~S.~Turner,
  Phys.\ Rev.\ D {\bf 36}, 2201 (1987).
  
\bibitem{Hall:2013eko} 
  L.~J.~Hall and Y.~Nomura,
  JHEP {\bf 1402}, 129 (2014)
  [arXiv:1312.6695 [hep-ph]].
  
\bibitem{Hall:2014vga} 
  L.~J.~Hall, Y.~Nomura and S.~Shirai,
  JHEP {\bf 1406}, 137 (2014)
  [arXiv:1403.8138 [hep-ph]].
  
\bibitem{Linde:1979ny} 
  A.~D.~Linde,
  Phys.\ Lett.\ B {\bf 92}, 119 (1980).

\bibitem{Lindner:1985uk} 
  M.~Lindner,
  Z.\ Phys.\ C {\bf 31}, 295 (1986).

\bibitem{Sher:1993mf} 
  M.~Sher,
  Phys.\ Lett.\ B {\bf 317}, 159 (1993)
  [Addendum-ibid.\ B {\bf 331}, 448 (1994)]
  [hep-ph/9307342].

\bibitem{Holthausen:2011aa} 
  M.~Holthausen, K.~S.~Lim and M.~Lindner,
  JHEP {\bf 1202}, 037 (2012)
  [arXiv:1112.2415 [hep-ph]].
  
\bibitem{EliasMiro:2011aa} 
  J.~Elias-Miro, J.~R.~Espinosa, G.~F.~Giudice, G.~Isidori, A.~Riotto and A.~Strumia,
  Phys.\ Lett.\ B {\bf 709}, 222 (2012)
  [arXiv:1112.3022 [hep-ph]].

\bibitem{Degrassi:2012ry} 
  G.~Degrassi, S.~Di Vita, J.~Elias-Miro, J.~R.~Espinosa, G.~F.~Giudice, G.~Isidori and A.~Strumia,
  JHEP {\bf 1208}, 098 (2012)
  [arXiv:1205.6497 [hep-ph]].
 
\bibitem{Buttazzo:2013uya} 
  D.~Buttazzo, G.~Degrassi, P.~P.~Giardino, G.~F.~Giudice, F.~Sala, A.~Salvio and A.~Strumia,
  arXiv:1307.3536 [hep-ph].

 \bibitem{Feldstein:2006ce} 
  B.~Feldstein, L.~J.~Hall and T.~Watari,
  Phys.\ Rev.\ D {\bf 74}, 095011 (2006)
  [hep-ph/0608121].

\bibitem{D'Eramo:2014rna} 
  F.~D'Eramo, L.~J.~Hall and D.~Pappadopulo,
  JHEP {\bf 1411}, 108 (2014)
  [arXiv:1409.5123 [hep-ph]].
 
\bibitem{Coleman:1973jx} 
  S.~R.~Coleman and E.~J.~Weinberg,
  Phys.\ Rev.\ D {\bf 7}, 1888 (1973).
  
\bibitem{Giudice:1998xp} 
  G.~F.~Giudice, M.~A.~Luty, H.~Murayama and R.~Rattazzi,
  JHEP {\bf 9812}, 027 (1998)
  [hep-ph/9810442].

\bibitem{Gherghetta:1999sw} 
  T.~Gherghetta, G.~F.~Giudice and J.~D.~Wells,
  Nucl.\ Phys.\ B {\bf 559}, 27 (1999)
  [hep-ph/9904378].
  
\bibitem{Cheung:2011mg} 
  C.~Cheung, G.~Elor and L.~J.~Hall,
  Phys.\ Rev.\ D {\bf 85}, 015008 (2012)
  [arXiv:1104.0692 [hep-ph]].

\bibitem{Tegmark:2005dy}
M.~Tegmark, A.~Aguirre, M.~J.~Rees and F.~Wilczek,
Phys.\ Rev.\  D {\bf 73}, 023505 (2006)
[arXiv:astro-ph/0511774].

\bibitem{Bae:2008ue} 
  K.~J.~Bae, J.~H.~Huh and J.~E.~Kim,
  JCAP {\bf 0809}, 005 (2008)
  [arXiv:0806.0497 [hep-ph]].

\bibitem{Freivogel:2008qc} 
  B.~Freivogel,
  JCAP {\bf 1003}, 021 (2010)
  [arXiv:0810.0703 [hep-th]].
  
\bibitem{Bousso:2013rda} 
  R.~Bousso and L.~Hall,
  Phys.\ Rev.\ D {\bf 88}, 063503 (2013)
  [arXiv:1304.6407 [hep-th]].

\bibitem{Mahbubani:2005pt} 
  R.~Mahbubani and L.~Senatore,
  Phys.\ Rev.\ D {\bf 73}, 043510 (2006)
  [hep-ph/0510064].

\bibitem{D'Eramo:2007ga} 
  F.~D'Eramo,
  Phys.\ Rev.\ D {\bf 76}, 083522 (2007)
  [arXiv:0705.4493 [hep-ph]].
  
\bibitem{Enberg:2007rp} 
  R.~Enberg, P.~J.~Fox, L.~J.~Hall, A.~Y.~Papaioannou and M.~Papucci,
  JHEP {\bf 0711}, 014 (2007)
  [arXiv:0706.0918 [hep-ph]].

\bibitem{Cohen:2011ec} 
  T.~Cohen, J.~Kearney, A.~Pierce and D.~Tucker-Smith,
  Phys.\ Rev.\ D {\bf 85}, 075003 (2012)
  [arXiv:1109.2604 [hep-ph]].



  





  

  
  
    

  
\bibitem{Salam:1974jj} 
  A.~Salam and J.~A.~Strathdee,
  Phys.\ Rev.\ D {\bf 11}, 1521 (1975).
  
\bibitem{Grisaru:1979wc} 
  M.~T.~Grisaru, W.~Siegel and M.~Rocek,
  Nucl.\ Phys.\ B {\bf 159}, 429 (1979).
  
\bibitem{Seiberg:1993vc} 
  N.~Seiberg,
  Phys.\ Lett.\ B {\bf 318}, 469 (1993)
  [hep-ph/9309335].
  
\bibitem{Machacek:1983fi} 
  M.~E.~Machacek and M.~T.~Vaughn,
  Nucl.\ Phys.\ B {\bf 236}, 221 (1984).
  
\bibitem{Ellwanger:2009dp} 
  U.~Ellwanger, C.~Hugonie and A.~M.~Teixeira,
  ``The Next-to-Minimal Supersymmetric Standard Model,''
  Phys.\ Rept.\  {\bf 496}, 1 (2010)
 [arXiv:0910.1785 [hep-ph]].

  
 \end{thebibliography}
\end{document}